\documentclass[%
 reprint,
superscriptaddress,
 amsmath,amssymb,
 aps,
]{revtex4-2}

\usepackage{graphicx}% Include figure files
\usepackage{dcolumn}% Align table columns on decimal point
\usepackage{bm}% bold math
\usepackage{xargs}                      % Use more than one optional parameter in a new commands
\usepackage[pdftex,dvipsnames]{xcolor}  % Coloured text etc.
\usepackage{xr} % reference figures from supplementary
\begin{document}

\newcommand{\chlamy}{\textit{C. reinhardtii}}
\newcommand{\zcm}{z_{\rm cm}}
\newcommand{\mmol}{\mu \rm mol\cdot m^{-2}\cdot s^{-1}}
\newcommand{\sbound}{s^{\star}}
\newcommand{\sfree}{s}
\newcommand{\stot}{s_{\rm tot}}
\newcommand{\microns}{\; \mu \rm m}
\newcommand{\etal}{\textit{et al.}}

\newcommand{\gabriel}[1]{{\textcolor{Black}{#1}}}
\newcommand{\mojtaba}[1]{{\textcolor{Black}{#1}}}
\newcommand{\taha}[1]{{\textcolor{Black}{#1}}}
\newcommand{\reviewerone}[1]{{\textcolor{Black}{#1}}}
\newcommand{\reviewertwo}[1]{{\textcolor{Black}{#1}}}
\newcommand{\reviewerthree}[1]{{\textcolor{Black}{#1}}}

\newcommand{\ladhyx}{Laboratoire  d'Hydrodynamique (LadHyX), CNRS, Ecole Polytechnique, Institut Polytechnique de Paris, 91120 Palaiseau, France}
\newcommand{\fast}{Université Paris-Saclay, CNRS, FAST, 91405 Orsay, France}

%\title{Memory effects in the phototactic behavior of \textit{Chlamydomonas reinhardtii} }

\title{Short-term memory effects in the phototactic behavior of microalgae}

\author{Taha Laroussi}
\affiliation{\ladhyx}

\author{Mojtaba Jarrahi}%
\affiliation{\fast}

\author{Gabriel Amselem}
\affiliation{\ladhyx}
\email{gabriel.amselem@polytechnique.edu}

\date{\today}

\begin{abstract}
Phototaxis, the directed motion in response to a light stimulus, is crucial for motile microorganisms that rely on photosynthesis, such as the unicellular microalga \textit{Chlamydomonas reinhardtii}. It is well known that microalgae adapt to ambient light stimuli. On time scales of several dozen minutes, when stimulated long enough, the response of the microalga evolves as if the light intensity were decreasing~[Mayer, \textit{Nature} (1968)]. Here, we show experimentally that microalgae also have a short-term memory, on the time scale of a couple of minutes, which is the opposite of adaptation. At these short time scales, when stimulated consecutively, the response of \chlamy\ evolves as if the light intensity were increasing. Our experimental results are rationalized by the introduction of a simplified model of phototaxis. Memory comes from the interplay between an internal biochemical time scale and the time scale of the stimulus; as such, these memory effects are likely to be widespread in phototactic microorganisms.
\end{abstract}

\maketitle

\section{Introduction}

Phototaxis, the directed motion of organisms in response to a light stimulus, is widespread both in prokaryotes and single-cell eukaryotes~\cite{jekely2009evolution}. One of the cellular models for eukaryotic phototaxis is the microalga \textit{Chlamydomonas reinhardtii}, which responds to blue-green light~\cite{foster1980light,hegemann2009sensory}. When light hits the eyespot of the microalga, it induces photocurrents, whose amplitude depend on the light intensity. These photocurrents then induce flagellar currents which change the beating pattern of the flagella, leading to reorientation and eventually phototaxis~\cite{hegemann2009sensory,ruffer1991flagellar}.

At low light intensities, wild-type \chlamy\ cells swim towards the light, while they swim away from the light at high light intensities. This corresponds to positive and negative phototaxis, respectively~\cite{feinleib1971relationship}. What sets the change in phototactic behavior? The answer to this question is not fully settled~\cite{hegemann2009sensory}.

Several biochemical parameters were found to affect the sign of phototaxis of \chlamy , such as the amount of calcium ions in the surrounding medium~\cite{nultsch1986effects,morel-laurens1987calcium,dolle1987role,witman1993chlamydomonas}, photosynthetic activity of the microalgae \reviewertwo{~\cite{takahashi1993photosynthesis, arrieta2017phototaxis}}, the amount of intracellular reactive oxygen species \reviewertwo{~\cite{wakabayashi2011reduction, erickson2015light}}, or the phosphorylation of channelrhodopsin-1, one of the photoreceptors of \chlamy~\cite{bohm2019channelrhodopsin}. 

It is also known that the history of the alga plays a role in its phototactic response: like many unicellular organisms ~\cite{macnab1972gradient, berg1975transient, tawada1972responses, van-haastert1983sensory}, \chlamy\ adapt to their environment. \reviewertwo{When stimulated repeatedly with the same stimulus at low light intensities,  \chlamy\ exhibits a  positive phototactic behavior which is less and  less pronounced. Such an acclimation occurs with a characteristic time scale of $\approx 30\; \rm s$, similar to the time scale of adaptation of the photosynthetic apparatus~\cite{arrieta2017phototaxis}.} Repeated stimuli can also lead to changes in the phototactic sign of a population. 
A naive cell population, kept in the dark, exhibits negative phototaxis in response to an intense light stimulus; the same cell population, exposed to the same intense light stimulus, undergoes positive phototaxis when it has been previously exposed to light for a couple dozens minutes~\cite{mayer1968chlamydomonas}. This qualitative change in the phototactic behavior of a population depending on the history of irradiation is consistent with results obtained at the single-cell level by R\"{u}ffer and Nultsch~\cite{ruffer1991flagellar}, who monitored the change in beating of the two flagella of \chlamy\ in response to increasing and decreasing light-stimuli. Such a history-dependent change in the phototactic response occurs on long time scales, of the order of a couple dozen minutes.

Here, we show experimentally that the change in phototactic sign also depends on the recent history of the cell, where the time scale is of the order of a couple minutes. At these time scales, the algae exhibit a behavior that is the exact opposite of the long-term adaptation: they integrate consecutive signals over time. When subjected to two consecutive, closely spaced identical stimuli, an alga essentially adds up the second stimulus to the first one, and acts as if the second stimulus were of higher intensity than the first one. Such a signal integration has, to the best of our knowledge, never been observed in the phototactic response of microalgae. Our results are rationalized by the introduction of a simplified model of phototaxis. The memory emerges in the model from the interplay between two time scales, the time between successive stimuli and the relaxation time of an inner biochemical process. Since the model is generic, similar short-term memory effects are likely to be widespread in other organisms that experience phototaxis.

\section{Methods}
\subsection{Culture preparation}

\chlamy\ strain CC-125 (Chlamydomonas Resource Center, University of Minnesota, MN, USA) were cultured on a solid medium prepared with Tris-Acetate-Phosphate (Gibco{\texttrademark} TAP, ThermoFischer Scientific, France) and agar every 4 weeks, to keep the strain motile, responsive to light and prevent as much as possible the cells from sticking to the walls of the devices. For experiments, colonies of \chlamy\ were picked from solid cultures and recultured in liquid TAP. The cultures were placed in an agitating incubator at $176\; \rm rpm$, under a day/light cycle of 14h/10h with an illuminance of $40\; \mmol$ and at a fixed temperature of $22^{\circ}\rm C$. The cells reached maximum motile cell density within 3 days \cite{harris2009chlamydomonas}, after which a solution of swimming \chlamy\ was obtained. 

\reviewerone{It is known that the phototactic response of microalgae may be regulated by its inner biological circadian clock through the day \cite{bruce1970biological}. For experiments, the culture was systematically taken 4 hours after the beginning of the ``day'', and underwent three centrifugation steps  to concentrate the algal solution and get rid of low-motility algae, dead algae, and cellular debris. Algae were then left to rest for one hour in the dark before the beginning of an experiment, ensuring all algae were motile, swimming at an average speed $\approx 60 \; \rm \mu m\cdot s^{-1}$. A detailed protocol can be found in the Supplementary Material.}

\subsection{Experimental setup}

An array of shallow cylindrical wells was made in  polydimethylsiloxane (PDMS, Dow-Corning Sylgard 184),  using standard soft lithography techniques~\cite{xia1998soft}. The wells had a depth of $32\microns$ and  diameters ranging from $200\microns$ to $500\microns$. The PDMS was rendered hydrophilic by plasma cleaning. Then, a drop of $\approx 10 \; \rm \mu L$ of the algal solution was deposited onto the PDMS. The device was closed with a  plasma-activated glass slide by applying gentle pressure. 

\begin{figure}[h!]
    \centering
    \includegraphics[width=0.5\textwidth]{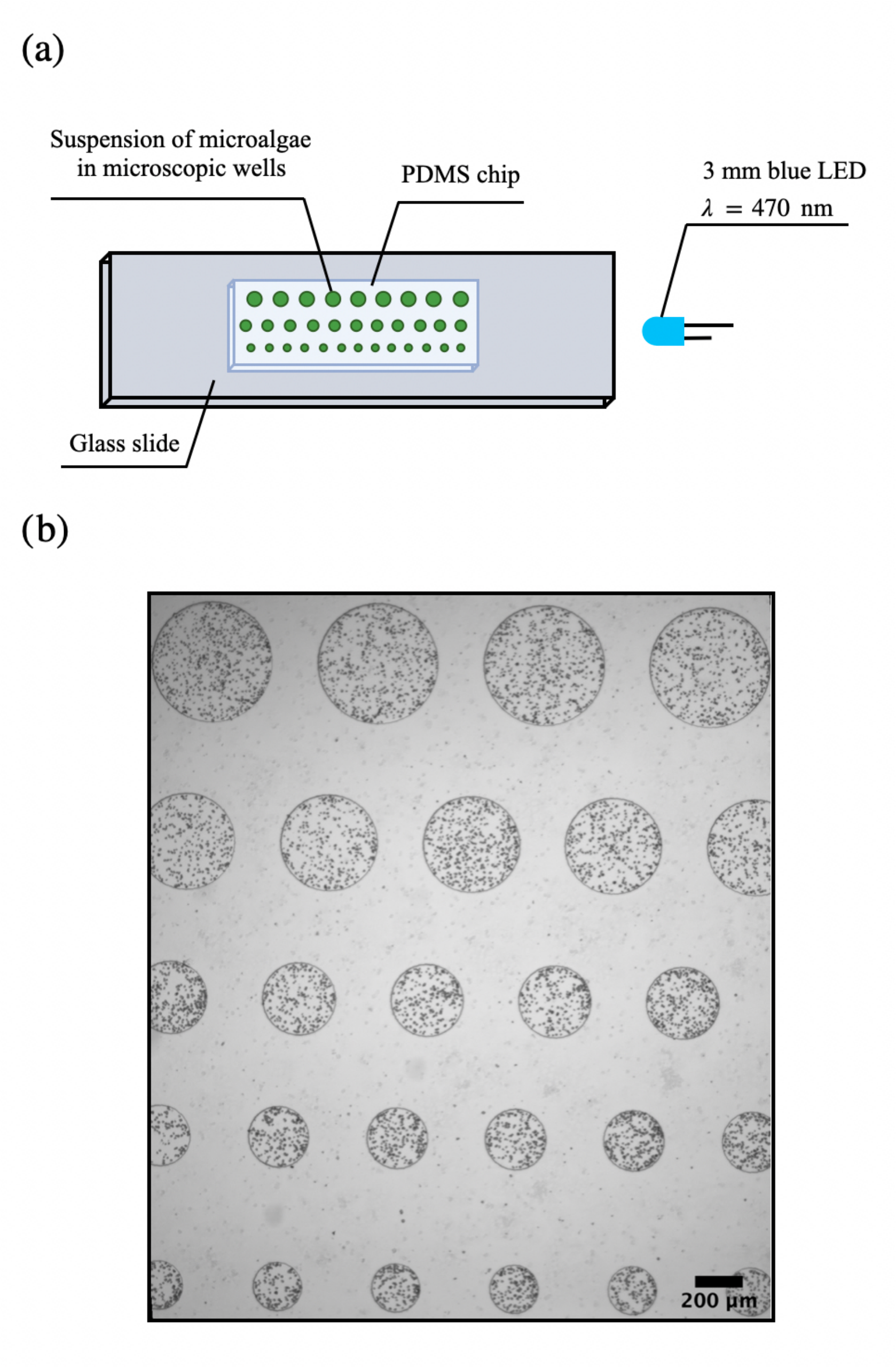}
    \caption{Experimental setup. %of an algae population trapped in quasi 2D cylindrical wells.
    (a) Schematic drawing of a PDMS chip with shallow cylindrical wells of varying diameters. The algal solution is deposited in the wells, which are then sealed with a glass slide. The chip is exposed to blue light from the side. (b) Bright-field microscopy image of an array of 23 cylindrical wells of different diameters and identical algae concentrations. Cell bodies are visible as black dots inside the wells.}
    \label{fig:setup_fig}
\end{figure}

The trapped algae were observed under a Nikon TI microscope, using a 4X objective. Images were recorded at 10 fps with a CMOS camera (Hamamatsu ORCA-Flash4.0 LT, Hamamatsu Photonics, France). A long-pass red filter with a $645\; \rm nm$ cut-on wavelength (Newport RG645) was placed between the microscope's light source and the microwells to \reviewerone{monitor the algae's behavior without triggering a phototactic response~\cite{sineshchekov2002two,leptos2009dynamics}}. A 3~mm blue LED \reviewerone{with a peak intensity at $\lambda = 486\; \rm nm$ and a full width at half maximum $\approx 29\; \rm nm$} %and a light intensity of $7000\; \rm mcd$ 
(Plan\`{e}te Leds, France) was placed on the microscope's plate, approximately $2\; \rm cm$ away from the microfluidic chip, on the same plane, to stimulate the algae, see Fig.~\ref{fig:setup_fig}a. 
The blue light intensity was tuned by varying the voltage applied to the LED. \reviewerone{In between recordings, algae were kept in the dark.}

\reviewerone{The average density of algae in the wells was controlled by adjusting the concentration of the solution of algae. At the beginning of the experiments, before applying any light stimulus, the algae swam randomly and were distributed homogeneously in the wells. The algae were identified by binarizing the images (see Supp. Fig.~1), and the total area they occupied in each well was calculated. This area was renormalized by the well area, leading to the definition of the projected area fraction $\phi$, which  was used as a proxy for the algae concentration. The relative error on the projected area fraction $\phi$ was estimated to be of the order of 10\% (see Supp. Fig.~2), mainly because of variations in the thresholding outcome from image to image. We verified that, at cell densities used in the article, the projected area fraction $\phi$ was a good proxy for the cell concentration, see Supp. Fig.~2}

\begin{figure*}[!ht]
  \centering
  \includegraphics[width=1\linewidth]{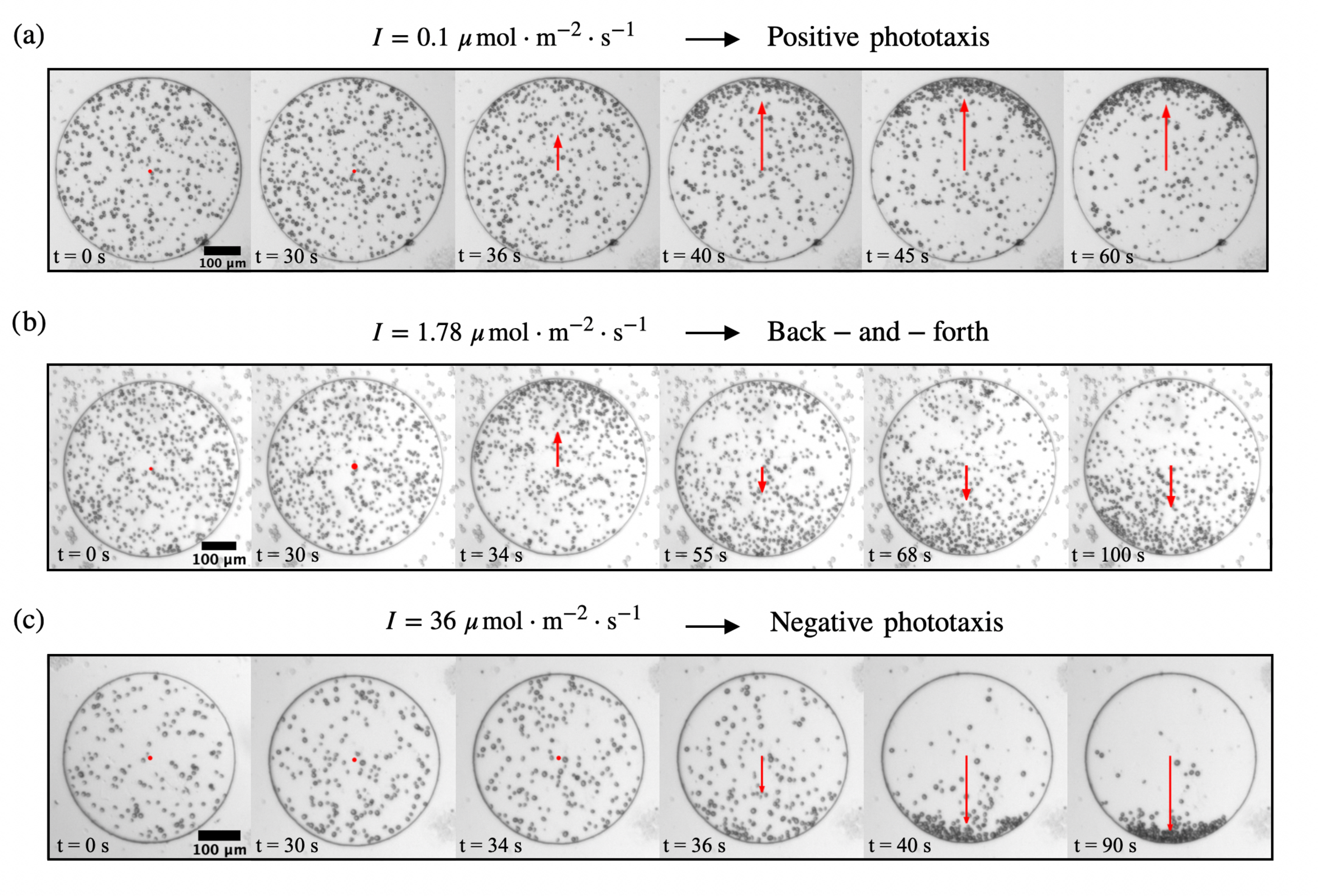}%
    \centering
    \caption{Different phototactic behaviors of \chlamy\ enclosed in a well. The blue light stimulus comes from the upper side of the wells and is turned on at $t=30\; \rm s$. (a) When exposed to a low intensity $I = 0.1\; \mmol$, the algae show positive phototaxis. (b) At intermediate intensities $I = 1.78 \; \mmol$, the algae show first positive phototaxis, then negative phototaxis. (c) At a high intensity $I = 36 \; \mmol$, the algae show negative phototaxis. The center of mass of the population is represented with a red arrow pointing from the center of the well.}
    \label{fig:snapshots}
\end{figure*}

\subsection{Measuring the local light intensity}
\label{sec:light_intensity_calculation}

\reviewerone{To estimate the flux of photons reaching the algae, we proceeded in two steps. First, before the start of an experiment, an image of the wells was taken using the blue LED as the sole source of light.  The gray value levels recorded by the camera could then be converted to a flux of photons reaching the camera. This corresponds to the blue light scattered by the PDMS, as the LED is on a perpendicular axis compared to the optical axis of the microscope. In a second step, we related this scattered light to the photon flux reaching the algae using a calibration curve, see Supp. Fig.~3. The detailed protocol can be found in the Supplementary Material.}

\section{Results}

\subsection{First experimental results}

\begin{figure*}[ht!]
  \centering
  \includegraphics[width=1\linewidth]{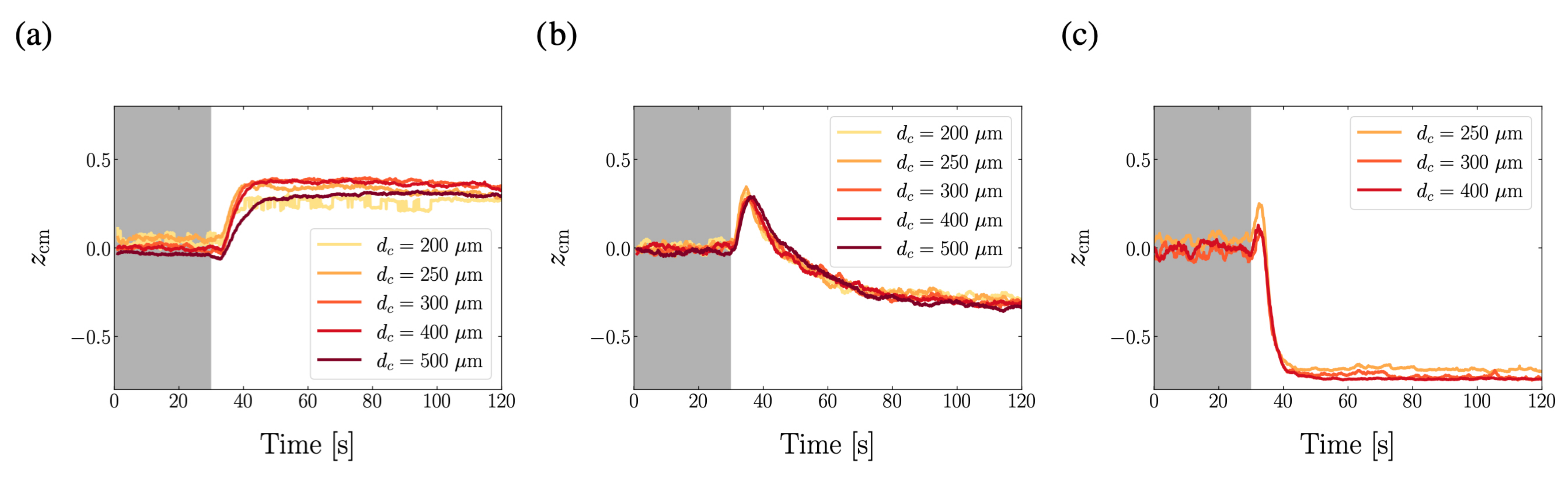}%
    \centering
    \caption{The diameter of the well $d_c$ does not influence the phototactic behavior of the algae. The light stimulus is turned on at $t=30\; \rm s$ and the position of the center of mass $\zcm$ is tracked over time.  (a) When exposed to an intensity $I = 0.38 \; \mmol $, the algae show positive phototaxis. (b) At intermediate intensities $I = 2.4 \; \mmol$, the algae show first positive phototaxis, then negative phototaxis. (c) At a high intensity $I = 28 \; \mmol$, the algae show negative phototaxis. Each curve is an average over 4 to 7 experiments.}
    \label{fig:diameter}
\end{figure*}

The experiments consist in exposing a population of \chlamy\ to a directional source of light. 
When a population of microalgae is exposed to a light intensity lower than $I_{\rm low}\approx 1.3\; \mmol$, the algae systematically exhibit positive phototaxis and swim towards the light source. Within $5$ to $8$ seconds of stimulation, most of the algae have accumulated at the boundary facing the light, see snapshots in Fig.~\ref{fig:snapshots}a. The algae remain accumulated at the boundary during at least one minute, time after which the imaging is stopped.  When the light intensity is higher than $I_{\rm high}\approx 8.6\; \mmol$, the algae always exhibit negative phototaxis and accumulate at the boundary opposing the light source, see snapshots in Fig.~\ref{fig:snapshots}c. These  results are well known in the literature although the precise value of the threshold intensities  varies from experiment to experiment, see e.g.~\cite{ueki2016eyespot,ramamonjy2022nonlinear}. \\
In between these two light intensities, there is a transition regime, where the phototactic behavior is hard to reproduce, despite keeping all experimental parameters identical: the microalgae sometimes exhibit a transient positive phototaxis followed by a negative phototaxis (see Fig.~\ref{fig:snapshots}b),  but can also display purely positive phototaxis, or purely negative phototaxis (see later in text). 
The aim of our work is to better understand the parameters responsible for this variety of algal behaviors in the transition regime. We investigate the effects of well diameter, algae concentration, light intensity, and finally history of the algae, on the phototactic behavior.

\subsection{Quantification of the phototactic behavior}

To quantify the phototactic behavior of a microalgae population constrained in its well, we binarize the experimental images using a simple Otsu threshold on pixel intensities \reviewertwo{\cite{otsu1979threshold}}. This leads to images where the algae are white, on a black background (see Supp. Fig.~1). In each image, we then calculate the center of mass $\zcm$ of the white pixels, corresponding to the center of mass of the population. The position of the center of mass is tracked over time, and eventually normalized to the radius of the well. A value of the center of mass $\zcm = 1$ (resp. $\zcm = -1$) corresponds to all algae accumulating at the boundary of the well facing (resp. opposite from) the light.

In one experimental run, the phototactic response of 15 -- 30 wells containing microalgae is quantified. The field of view contains wells with at least 3 different diameters, and there are at least 4 wells of each diameter, see Fig.~\ref{fig:setup_fig}b. We start with an experiment where the algae are at the same concentration in all wells, and do not find any impact of the well diameter on the phototactic \reviewerone{sign} of populations of \chlamy: for all diameters, in one given experiment, the algae exhibit the same behavior, as shown by the evolution of the center of mass of the population, see Fig.~\ref{fig:diameter}. Note that the center of mass never reaches $\pm 1$. This is mainly due to a fraction of the algae not responding to light, \reviewerone{(see Supp. Fig.~4 and~5). Rescaling the center of mass by the fraction of non-responding algae leads to qualitatively similar results, with rescaled centers of mass that reach positions closer to $\pm 1$,  see Supp. Fig.~6.} A second-order effect is that algae take space, and so the center of mass of the population can never reach $\pm 1$.

The time scales of accumulation were measured manually. Positive phototaxis leads to an accumulation on the side of the light source within $t_{\rm pos}\approx 6.3 \pm 0.6 \; \rm s$ (mean $\pm$ std. deviation over all experiments) after the stimulus is turned on. The time scale for negative phototaxis is slightly longer: $t_{\rm neg}\approx 10.2 \pm 1.9 \; \rm s$. In the case where there is a back-and-forth between positive and negative phototaxis, the first accumulation occurs within $t_{\rm pos}\approx 5.1 \pm 0.3 \; \rm s$ while the second accumulation is within  $t_{\rm neg}\approx 36.3 \pm 6.9 \; \rm s$. \reviewerone{Note that during one experimental run, the time scale of accumulation depends on the well diameter, with differences of a couple seconds between the smallest and largest wells, corresponding to the time needed to swim to a closer or further boundary. Over multiple experimental runs however, the variability due to the different diameters is blurred by the inter-experimental variability. In all cases, we have that negative phototaxis  occurs on a time scale longer than positive phototaxis. }

\subsection{Effect of cell concentration}

\begin{figure*}[htb!]
  \centering
  \includegraphics[width=1\linewidth]{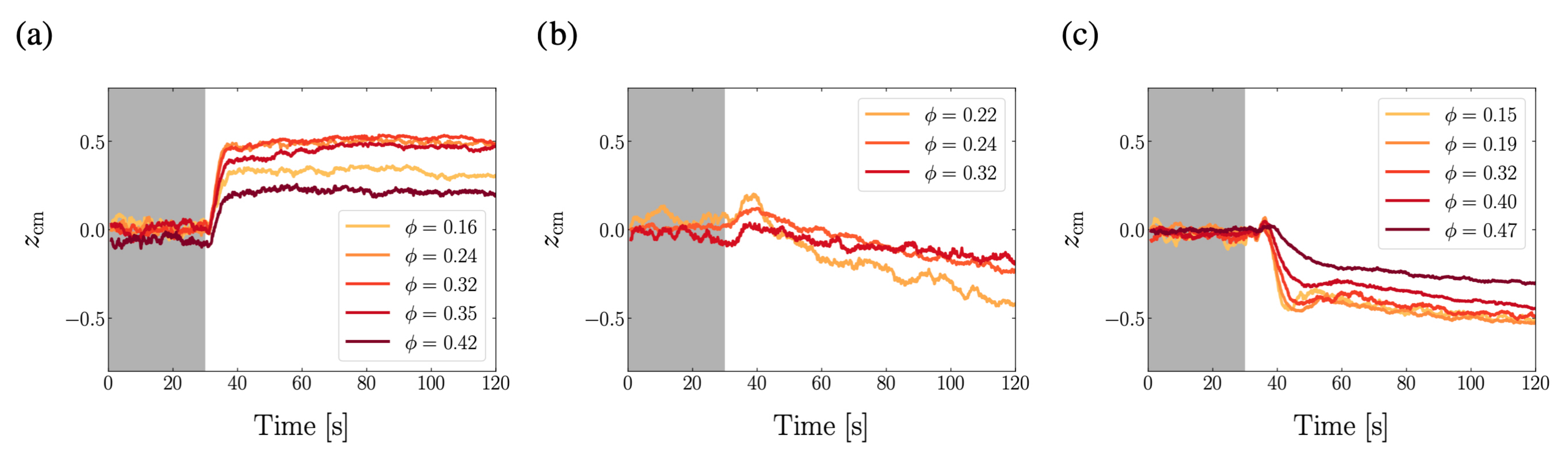}%
    \centering
    \caption{The concentration $\phi$ of algae in the well does not influence the phototactic behavior. The  light stimulus is turned on at $t=30\; \rm s$ and the position of the center of mass $\zcm$ is tracked over time. (a) When exposed to an intensity $I = 1.18 \; \mmol $, the algae show positive phototaxi (b) At intermediate intensities $I = 1.38 \; \mmol $, the algae show first positive phototaxis, then negative phototaxis. (c) At a high intensity $I = 17.6 \; \mmol $, the algae show negative phototaxis.
    %Each curve is an average over 1 to 4 experiments.
    }
    \label{fig:concentration}
\end{figure*}

\reviewerthree{It is known that the cell density plays a role in the response of \chlamy\ to light stimuli, and that increasing the cell density  leads to an overall more directional phototactic response~\cite{choudhary2019reentrant,furlan2012origin}. It is also known that algae can shield other algae from light, due to light absorption by the cell body~\cite{eisenmann2024collective}.} One possibility for the alternating phototactic behavior would then be to invoke screening of the light by the algae. We would  expect that at high algae concentration, the algae closest to the light source screen the light intensity, so that the algae further away from the light source see a dimmer light. This could result in a seemingly biphasic behavior, with, at high light intensities, algae close to the light source showing  negative phototaxis, and algae further away from the light source exhibiting positive phototaxis. 

To test this hypothesis, we enclosed microalgae at concentrations $\phi = 0.008$ to $\phi = 0.6$ in the microwells, and monitored the evolution of their center of mass. The concentration $\phi$ is defined as the surface fraction of algae in a microwell, see Methods. For $\phi > 0.5$, the well is essentially packed with algae, which impedes their motion and so leads to the center of mass not moving even during the light stimulus, see Supp. Fig.~7. For $\phi \leq 0.5$, we do not see any effect of the concentration on the sign of phototaxis: the phototactic behavior depends solely on the light intensity, see Fig.~\ref{fig:concentration}. %The rescaled barycenter positions are shown in Supp. Fig.~XXX.
Furthermore, there is an effect of concentration on the position of the center of mass $\zcm$: when the concentration is higher, the algae take more space and the center of mass of the population gets closer to the center of the well, see Fig.~\ref{fig:concentration}a and c.

\begin{figure}[h!]
    \centering
    \includegraphics[width=0.5\textwidth]{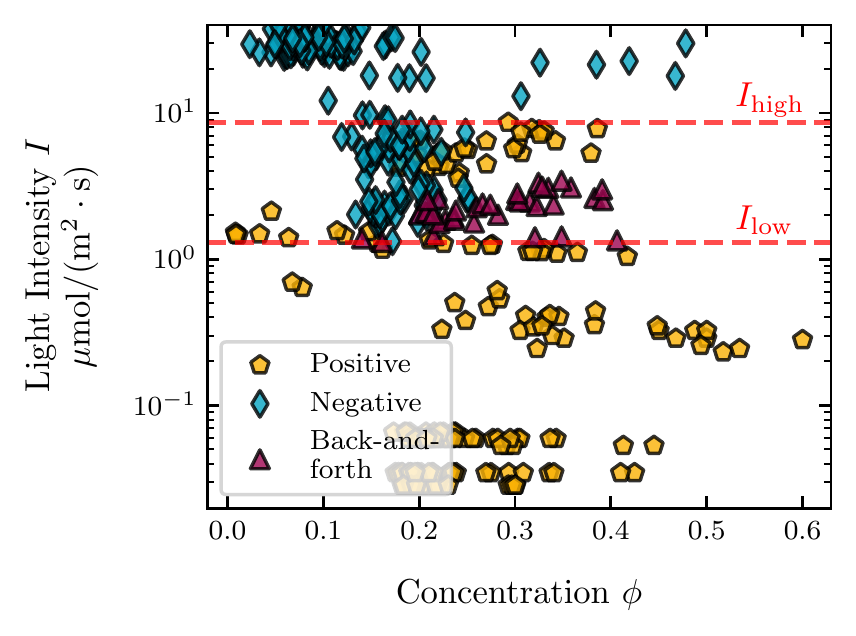}
    \caption{Phase diagram for the phototactic behavior of 367 populations of algae, as a function of light intensity and algae concentration.}
    \label{fig:phase_diagram}
\end{figure}

The experiments were reproduced over a range of light intensities and algae concentrations, in more than 300 wells. In each well, the behavior was quantified as positive phototaxis, negative phototaxis, or back-and-forth. The resulting phase diagram of phototactic behavior as a function of light intensity and algae concentration shows no clear influence of the algae concentration on the sign of phototaxis, see Fig.~\ref{fig:phase_diagram}.

\subsection{Effect of history}

\begin{figure*}[ht!]
  \centering
  \includegraphics[width=1\linewidth]{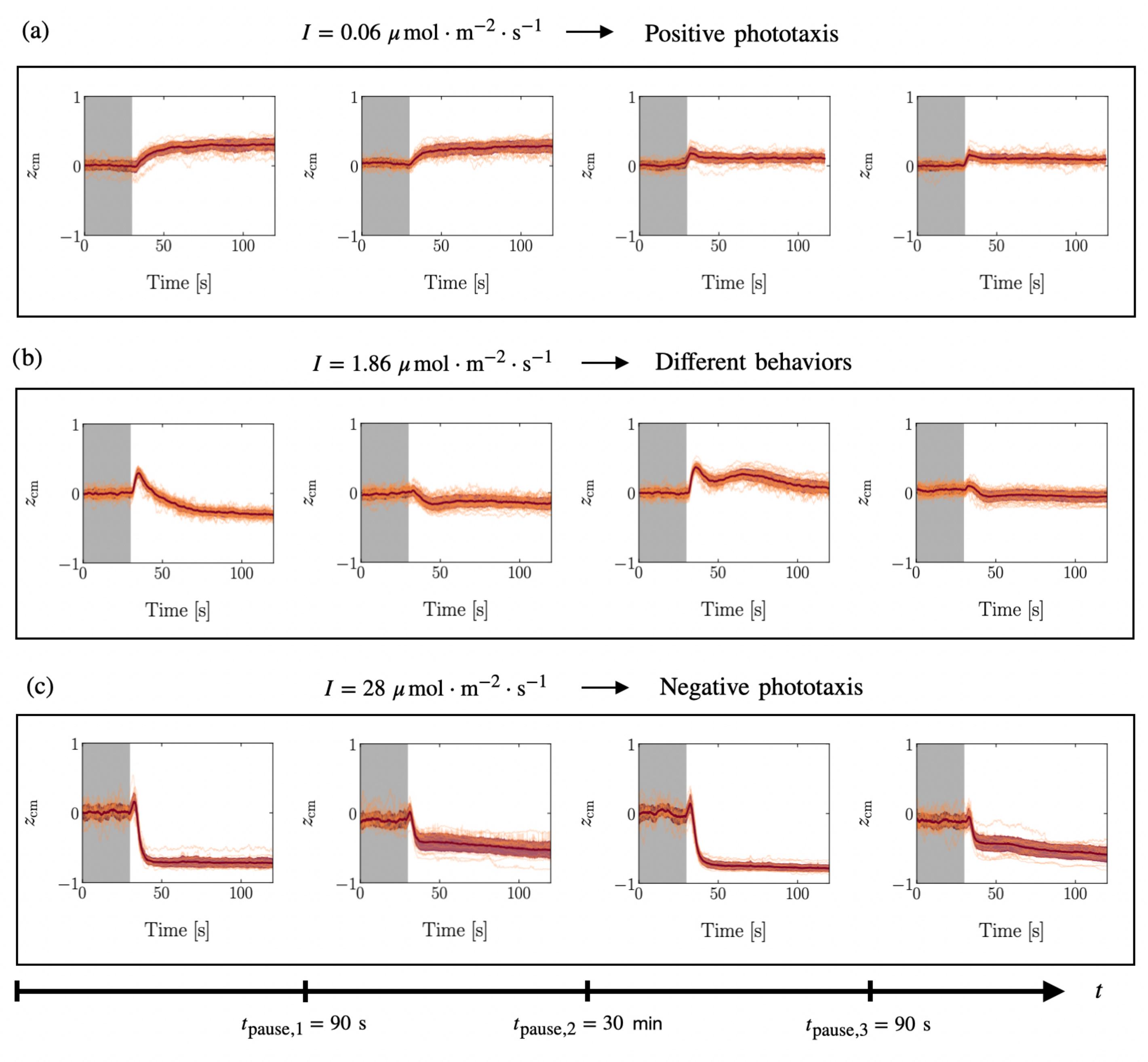}%
    \centering
    \caption{Memory effects on the phototactic response of algae. The light stimulus is turned on at $t=30\; \rm s$ and the position of the center of mass $\zcm$ is tracked over time. Four consecutive experiments are performed with varying rest times, where the  light is turned off in between. There is a 90 s pause between Exp. 1 and Exp. 2, then a 30 min pause between Exp. 2 and Exp. 3 and finally a 90 s pause between Exp. 3 and Exp. 4. (a) When exposed to a low intensity $I = 0.06 \; \mmol$, the algae show positive phototaxis in all experiments. The center of mass is averaged over 24 experiments. (b) At intermediate intensities $I = 1.86\; \mmol$, the algae show different phototactic behaviors in each experiment. The center of mass is averaged over 31 experiments. (c) At a high intensity $I = 28\; \mmol$, the algae show negative phototaxis in all experiments. The center of mass is averaged over 18 experiments. Each orange line represents the center of mass $\zcm$ of one experiment. The dark red solid line represents the average $\zcm$ for the set of experiments, and the shaded area in red gives the standard deviation around the mean.}
    \label{fig:time}
\end{figure*}

We finally investigated the effect of history on the behavior of the algae. To do so, a population of algae was kept in the dark for one hour and then exposed to a given light stimulus during 90 seconds. The light was then turned off during 90 seconds, after which the exact same stimulus was re-applied. \reviewerone{Algae were kept in the dark during all pauses.} At low light intensities, the qualitative behavior of algae does not change between the two stimuli: the algae always exhibit positive phototaxis, see the first two graphs in Fig.~\ref{fig:time}a. At high light intensities, the algae  always exhibit negative phototaxis, see the first two graphs in Fig.~\ref{fig:time}c. At intermediate light intensities however, the behavior of the algae changes between the two stimuli: in the first stimulus, the algae exhibit a back-and-forth motion, first towards the light and then away from it. During the second stimulus, the algae show negative phototaxis, see the first two graphs in Fig.~\ref{fig:time}b. 

To understand whether the change in behavior was permanent, the algae were allowed to rest in the dark for 30 minutes after the end of the second stimulus. Then, the same protocol was followed: a third stimulus, of the same intensity as the two previous ones, was applied. After 90 seconds in the dark, the algae were restimulated with a fourth, identical stimulus. In the case of low and high light intensity, the qualitative response of the algae was the same for the third and fourth stimuli as for the two first ones, see the last two graphs in Fig.~\ref{fig:time}a,c. At intermediate light intensities, the algae showed positive phototaxis during the third stimulus, and then qualitatively changed their behavior during the fourth stimulus to exhibit slightly negative phototaxis, see the last two graphs in Fig.~\ref{fig:time}b. This last response to light, averaged over 31
wells, is quite subtle. Indeed, after several stimuli, most of the algae stick to the wells, a likely effect of light stimulation~\cite{catalan2023light, kreis2018adhesion}. These stuck algae move very slowly by gliding~\cite{till2022motility}, see Supp. Fig.~8. The center of mass of the population then largely reflects the position of the stuck algae. The motile algae however do exhibit negative phototaxis, see Supp. Fig.~8. 

\reviewerthree{We also tested resting times of 5, 10 and 20 minutes between consecutive stimuli. No qualitative change of the phototactic behavior could be observed, see Supp. Fig.~9 and~10. The change in behavior after 90 seconds was however systematic, see Supp. Fig.~10.}

Note that the quantitative response of the algae, as measured by the position of the center of mass $\zcm$, differs between the four stimuli -- even in the case of low and high light intensities. For example, at low light intensities (positive phototaxis), the shift in the center of mass is identical for stimuli 1 and 2. The shift is less pronounced for stimuli 3 and 4, see Fig.~\ref{fig:time}a. In contrast, in the case of negative phototaxis, the shift of the center of mass is identical for stimuli 1 and 3, as well as for stimuli 2 and 4, see Fig.~\ref{fig:time}c: there, resting in the dark for 30 minutes seems to reset the phototactic behavior. 

What is important in our case however is the qualitative response: the algae can change phototactic sign at intermediate light intensities due to the stimuli they previously experienced. The change in behavior is not an adaptive response, which would make the algae show first negative phototaxis, then positive phototaxis in response to the same stimulus. It rather corresponds to an integration of the signal over time. This effect explains the variety of behaviors observed in the transition regime between positive and negative phototaxis. One cannot predict the phototactic response of an alga without knowing its history. 

\section{Model}

We now propose a simplified model to describe our experimental observations. We call  $I_0$ the light intensity within a well, and $S$ an intracellular biochemical species responsible for the sign of phototaxis. The total concentration of $S$ is conserved and is called $\stot$.  This species  exists in two states: active (with concentration $\sbound$) and inactive (with concentration $\sfree$), such that $\sbound+\sfree = \stot$.

We  assume the transition from inactive to active state depends on the light intensity $I_0$, while the reverse transition occurs at a constant rate $\tau^{-1}$: 
\begin{equation}
    \frac{d\sbound}{dt} = \gamma I_0 (\stot-\sbound) - \tau^{-1}\sbound,
    \label{eq:sEq}
\end{equation}
where $\gamma$ is the reaction rate at which $\sfree$ is converted into $\sbound$. When the concentration of active molecules $\sbound$ is large (resp. small), the algae undergo negative (resp. positive) phototaxis, which can be summarized as:
\begin{equation}
    \frac{dz}{dt} = -{\rm sign}(\sbound - s_T)v_0,
    \label{eq:xEq}
\end{equation}
where $z$ is the position of an alga, $s_T$ is the threshold concentration at which cells transition from positive to negative phototaxis, and $v_0$ is the characteristic speed of the alga. Equations~\eqref{eq:sEq} and~\eqref{eq:xEq} constitute a minimal model of phototaxis.

\begin{figure}[h!]
  \centering
  \includegraphics[width=1\linewidth]{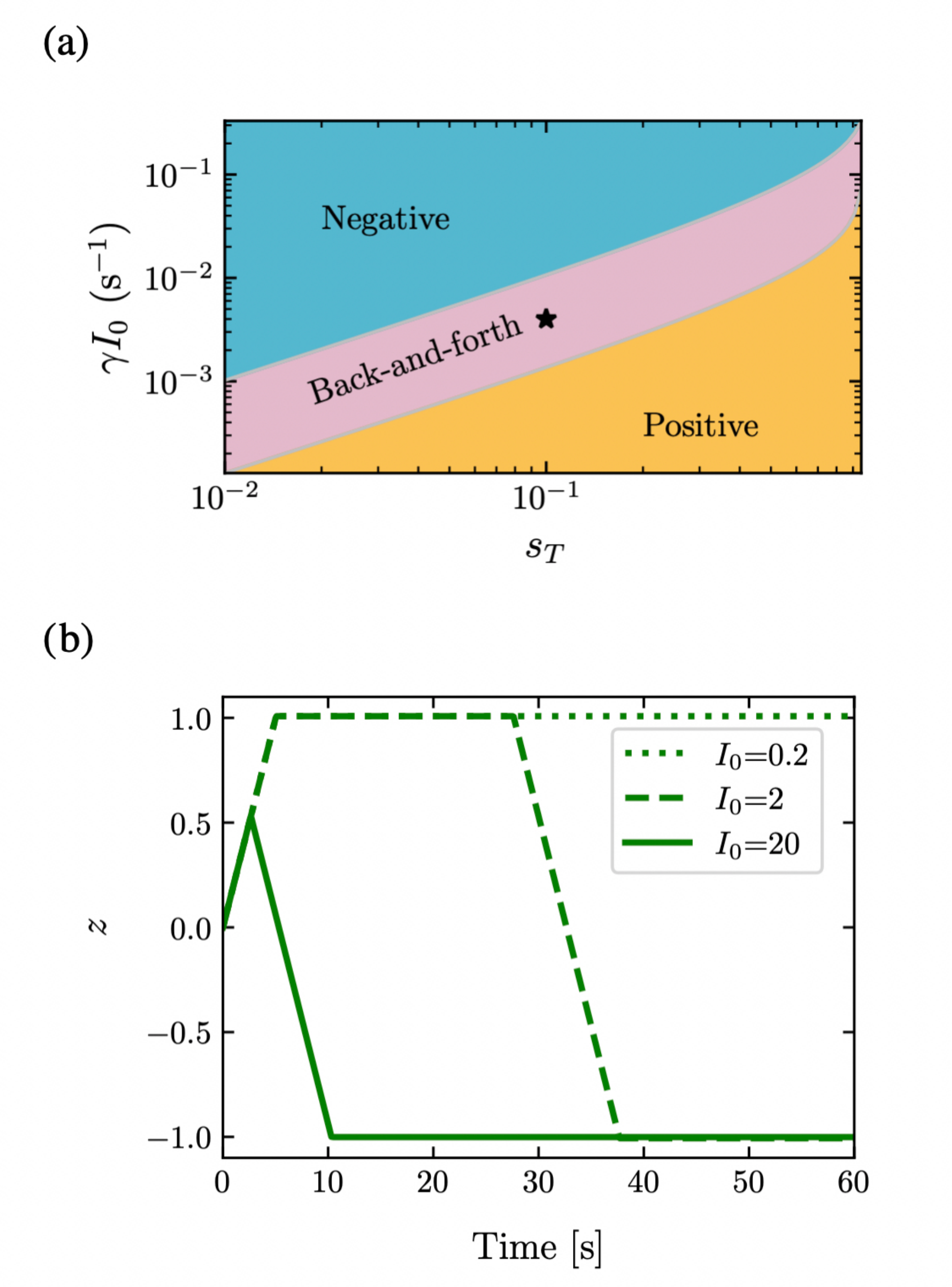}%
    \centering
    \caption{(a) Phase diagram of phototactic behavior as a function of the model parameters $\gamma I_0$ (activation rate of $s$), and the threshold $s_T$. \taha{The black star shows where the parameters $s_{T}$ and $\gamma$ are taken, ensuring a back-and-forth motion at $I_0 = 2 \; \mmol$ given the value of $\gamma = 2\times 10^{-3} \; \rm m^2\cdot \mu mol^{-1}$ used in the simulations.} (b) Simulation results for three different light intensities $I_0$. We recover positive phototaxis at $I_0 = 0.2\; \mmol$, back-and-forth motion at $I_0 = 2 \; \mmol$, and negative phototaxis at $I_0 = 20 \; \mmol$. Other simulation parameters: $s_{T} = 0.1$, $\stot = 1$, $v_0=100 \; \mu \rm m \cdot \rm s^{-1}$, $\tau = 300\;\rm  s$ and $\gamma = 2\times 10^{-3} \;\rm m^2\cdot \mu mol^{-1}$.}
    \label{fig:phaseSpaceModel}
\end{figure}

\reviewerthree{The model successfully reproduces the three behaviors observed experimentally. Without loss of generality, we normalize the concentration of $S$ so that $\stot = 1$. The amount of activated species $\sbound$  over time can be obtained from Eq.~\eqref{eq:sEq} (see Supp. Mat.), which enables to determine when the modelled cell undergoes positive phototaxis, back-and-forth motion, or negative phototaxis. We assume a cell is stimulated for 90~s at an intensity $I_0$, as in the experiments.  Positive phototactic behavior is obtained when $\sbound < s_T$ during the entire stimulus. In the experiments, negative phototaxis occurs in a time $t_{\rm neg} = 10\; \rm s$; in the simulations, this corresponds to $\sbound$ getting larger than $s_T$ in a time $t\leq t_{\rm neg} = 10\; \rm s$. In between these two conditions, we obtain a back-and-forth behavior. The obtained phototactic response is summarized as a function of the threshold $s_T$ and the value of $\gamma I_0$ in Fig.~\ref{fig:phaseSpaceModel}a, assuming a characteristic deactivation time $\tau=300\; \rm s$, consistent with the experiments. Varying the value of $\tau$ or $t_{\rm neg}$ does not change the phase space qualitatively.}

In the remaining part of this study, all simulations parameters are fixed except for the light intensity $I_0$. We place the algae in a well of radius $500\; \rm \mu m$. The position of the alga $z$ is then renormalized by this radius to obtain a number between -1 and 1. The light intensities are expressed, as in the experiment, as a flux  of photons in $\mmol$. We choose $\tau =300\; \rm s$, $\gamma=2\times 10^{-3} \; \rm m^2 \cdot \mu mol^{-1} $,  $s_T=0.1$, $\stot = 1$ and $v_0=100 \; \mu \rm m \cdot \rm s^{-1}$. \reviewerthree{The value of $\gamma$ is chosen to ensure that the minimal phototactic response occurs for $I_0 = 0.02\ \mmol$, as in the experiments (also see the Supplementary Material). At this value of $I_0$, we assume that only one molecule of $S$ is activated per second~\cite{ramamonjy2022nonlinear}. This determines $\gamma$. The choice of $\gamma$ then defines the choice of $s_T$, to obtain a back-and-forth motion at $I_0 = 2\; \mmol$, consistent with the experimental data.} Initially, the algae are naive ($\sbound = 0$) and $z = 0$.

At low light intensities ($I_0 = 0.2\; \mmol$), the amount of activated chemical species $\sbound$ is lower than the threshold $s_T$, and the resulting behavior is purely positive phototaxis. At intermediate light intensities ($I_0 = 2\; \mmol$), the usage rate $\gamma I_0$ and relaxation rate $\tau^{-1}$ \reviewerthree{are such that the threshold $s_T$ is crossed after $\approx 30\; s$}. Finally, at high light intensities ($I_0 = 20 \; \mmol$), all molecules are instantly converted to an active state,  and cannot relax back, resulting in negative phototaxis, see Fig.~\ref{fig:phaseSpaceModel}b. Note that, at high light intensities, the simulations show an initial small positive phototactic effect; this is similar to what is observed in experiments, see Fig.~\ref{fig:time}c, \reviewerone{and was previously reported in~\cite{catalan2023light}.}

Apart from the qualitative agreement between the different phototactic behaviours, the model also reproduces well the different experimental time scales. Positive phototaxis leads to accumulation within $t_{\rm pos}\approx 5 \; \rm s$, which depends solely on the swimming speed of the alga. Accumulation due to negative phototaxis occurs on a longer time scale, $t_{\rm neg}\approx 9\; \rm s$, due to the initial small positive phototactic effect. The back-and-forth motion leads to accumulation within $\approx 32 \; \rm s$, a consequence of the addition of three time scales: the time to swim to $z=1$, the time to produce  enough activated species to overcome the threshold $s_{T}$, and the time to eventually swim to $z=-1$.

We now simulate the application of successive stimuli of $90\; \rm s$, as in the experiments. Between two stimuli, the  alga position $z$ is reset to 0. Indeed, in the experiments, the algae repopulate homogeneously the well after one minute without light stimulus, and so the center of mass of the population goes back to 0 before the next experiment. At low ($I_0 = 0.2\; \mmol$) and high ($I_0 = 20\; \mmol$) light intensities, applying the same stimulus multiple times does not change the phototactic behavior, independently of the time between stimuli, see Fig.~\ref{fig:osci model/model_3_phototaxis_paused}a,c. 

At intermediate light intensity ($I_0 = 2\; \mmol$), we find that the phototactic behavior depends on the number of stimuli applied, their duration, and on the timing between stimuli. At the beginning of the first simulation, algae are in their naive state. When the stimulus is applied, the number of active molecules increases but is initially lower than the threshold $s_T$, which leads to positive phototaxis. When the amount of active molecules increases above the threshold value $s_T$, the phototactic behavior reverts during the experiment, and leads to the overall back-and-forth behavior. The light is then turned off for a duration $t=90\; \rm s$. After this rest time, the amount of active molecules has decayed but is still above $s_T$, so that when the algae are restimulated, they display purely negative phototaxis. We then remove the light stimulus for $30 \; \rm min$, allowing almost all active molecules to revert back to their inactive state. When the light is switched back on during the third experiment, the algae exhibit the back-and-forth motion: positive phototaxis at the beginning of the experiment, negative phototaxis once the threshold amount of active molecules is reached. Finally, the light is turned off once more for a duration $t=90\; \rm s$. After this rest time, the amount of active molecules is still above $s_T$, and the algae display purely negative phototaxis in the last simulation. Simulations results can be seen in Fig.~\ref{fig:osci model/model_3_phototaxis_paused}b, where the pause times are illustrated with gray-shaded areas. \reviewerone{The time evolution of $\sbound$ for all simulated experiments is shown in Supp. Fig.~11}. \reviewerthree{The phase diagram showing the phototactic behavior of algae after two consecutive stimuli of intensity $I_0$ and spaced by a time $t_{\rm pause}$ can be found in Supp. Fig.~12.}

A prediction of the model is that algae can switch behavior successively from positive phototaxis, to back-and-forth, to negative phototaxis before reverting back to positive phototaxis. Such is the case for three consecutive stimuli of $I_0=0.65\; \mmol$ spaced by $90\; \rm s$ and a pause of $30\; \rm min$ before the fourth stimulus, see Fig.~\ref{fig:osci model/model_3_phototaxis_paused}d. This switch through all phototactic behaviors requires a precise tuning of both the light intensity and the waiting times, which is experimentally tedious. We were however able to experimentally observe the switch from positive to back-and-forth behavior and from back-and-forth to negative behavior, see Supp. Fig.~13. 

\begin{figure*}[ht!]
  \centering
  \includegraphics[width=1\linewidth]{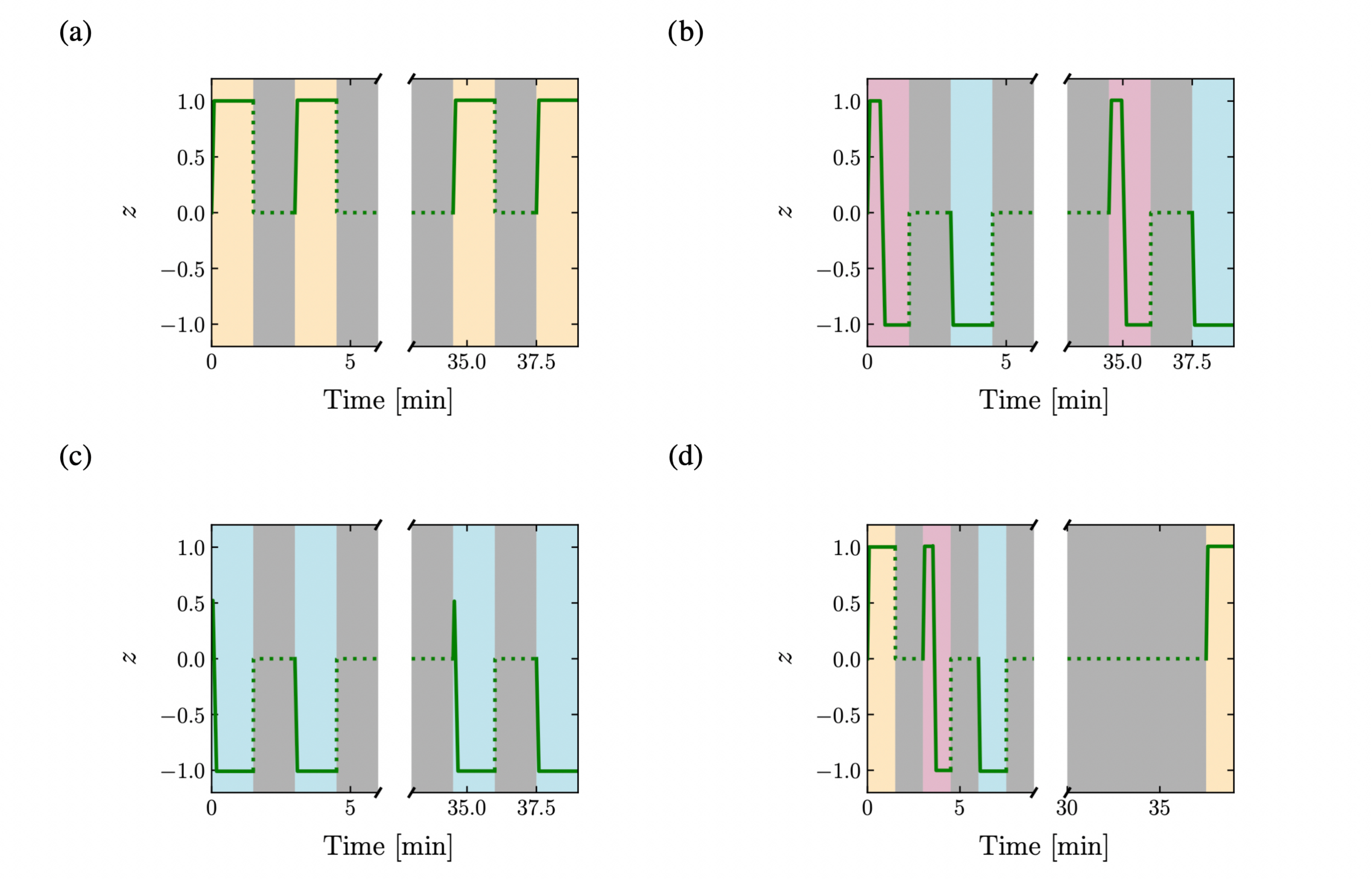}%
    \centering
    \caption{Simulation results for four consecutive experiments performed at a constant light intensity $I_0$, with different rest times in between. Each experiment lasts for $90$ s. Rest times are shown in shaded gray areas. ($t_{\rm{pause,1}}=90$ s, $t_{\rm{pause,2}}=30$ min and $t_{\rm{pause,3}}=90$ s for (a), (b) and (c)). (a) When exposed to a low light intensity $I_0 =0.2\; \mmol$, the alga always displays positive phototaxis. (b) At intermediate intensities $I_0 = 2\; \mmol$, the alga varies its phototactic behavior between each experiment. In the first experiment, the alga shows a back-and-forth motion. In the second, the alga shows only negative phototaxis. In the third experiment, the alga displays a back-and-forth motion again. Finally, in the last experiment, the behavior is negative phototaxis. (c) At a high intensity $I_0=20 \; \mmol$, the alga always shows negative phototaxis. (d) At low to intermediate intensities $I_0 = 0.65\; \mmol$ and with adjusted pause times ($t_{\rm{pause,1}}=90$ s, $t_{\rm{pause,2}}=90$ s and $t_{\rm{pause,3}}=30$ min) it is possible to successively go through positive phototaxis in the first experiment, back-and-forth motion in the second, negative phototaxis in the third and then revert back to positive phototaxis in the last experiment. Other simulation parameters: $s_{T} = 0.1$, $\stot = 1$, $v_0=100 \; \mu \rm m \cdot \rm s^{-1}$, $\tau = 300\; \rm s$ and $\gamma = 2 \times 10^{-3} \; \rm m^2\cdot  \mu mol^{-1}$.}
    \label{fig:osci model/model_3_phototaxis_paused}
\end{figure*}

\section{Discussion}
We have shown experimentally that the change in phototactic behavior of \chlamy\ can be due to its recent history of illumination. When exposed to two closely spaced, identical stimuli, the algae act as if they were time-integrating the stimuli. This is visible when the stimulus is close to that delimiting positive and negative phototaxis. Then, the algae show two different behaviors in response to the two consecutive stimuli: first a back-and-forth response, then a negative phototaxis. When the stimuli are sufficiently spaced apart, the algae have had time to relax to a basal state, and show the same qualitative response to the two stimuli, see Fig.~\ref{fig:time}. In our experiments, we find this time of complete relaxation to be of order 5 minutes. 

A simplified model of phototaxis explains this memory effect by the introduction of a time-scale of relaxation in the inner biochemistry of the cell. When the time-scale of relaxation is larger than the time between two consecutive stimuli, the alga integrates the signal and can display different behaviors in response to the two identical stimuli, see Fig.~\ref{fig:osci model/model_3_phototaxis_paused} \reviewerthree{and Supp. Fig. 11}. The time needed to return to the basal state should be of order $\tau$, which underlies our choice of $\tau = 5 \; \rm min$ for the relaxation time constant. %Our experimental data are consistent with an inner relaxation time $\tau$ in the range $\approx 5$ to 20 minutes. 

What could be the biochemical species corresponding to the intracellular species $S$ introduced in the model? It is very likely that our simplified model compounds the dynamics of multiple biochemical species, activated sequentially in a signalling cascade. In \chlamy , the photoreceptors localized in the eyespot whose activation triggers  phototaxis are channelrhodopsins, that become phosphorylated within milliseconds after the application of a light stimulus. A high level of phosphorylated channelrhodopsin-1 (ChR1) in the cell correlates with a change from positive to negative phototaxis~\cite{bohm2019channelrhodopsin}, much like a high level of activated $S$ corresponds to negative phototaxis in our model. Moreover, the time scale of dephosphorylation of ChR1 in \chlamy\ has been shown to be of the order of $\approx 5$ to 15 minutes~\cite{bohm2019channelrhodopsin}, which is compatible with our value of $\tau = 5\; \rm min$. A possibility is therefore to identify $S$ with ChR1. \reviewerone{Note that a second channelrhodospin, ChR2, is involved in sensing light at low intensities~\cite{sineshchekov2002two}, which is not considered in our model}. \reviewertwo{Another possibility is to think of the production of reactive oxygen species (ROS), which are produced in excess when the cell receives more light energy than it can use for photosynthesis~\cite{erickson2015light}. ROS can harm the photosynthetic apparatus through photo-oxidative damage ~\cite{niyogi1999photoprotection, li2022chloroplast,kottuparambil2012uv,erickson2015light}, and their concentration influences the sign of phototaxis~\cite{nakajima2023basis,wakabayashi2011reduction}. \chlamy\ employs various regulatory and protective mechanisms to reduce photo-oxidative damage ~\cite{niyogi1999photoprotection, witman1993chlamydomonas}, with different time scales~\cite{erickson2015light}. One of the mechanisms, zeaxanthin-dependent non-photochemical quenching of excited chlorophyll, has a time scale of several minutes, which could be compatible with our observations~\cite{niyogi1997chlamydomonas}.}

A more physical model of phototaxis was recently introduced by Leptos \etal\ to describe the response of \chlamy\ to light~\cite{leptos2023phototaxis}. In their model, Leptos \etal\ take into account the rotation of the microalga around its axis while swimming, and the resulting oscillating light intensity seen by the eyespot. The consecutive step-up and step-down stimuli sensed by the eyespot during one body rotation lead to a realignment of the trajectory towards the light source. Our model averages out the body rotation of the microalgae and describes the  behavior of the microorganism at a more coarse-grained time-scale. Another difference is that the model of Leptos \etal\ does not take into account the fact that phototaxis changes sign when the light intensity increases. This change of phototactic behavior requires the introduction of a threshold value in the model, below (resp. above) which phototaxis is positive (resp. negative). Such a threshold is incorporated in our simplified model. 

Note that the threshold in the model, introduced as a simple means to obtain a change in behavior between positive and negative phototaxis at different light intensities, leads to the back-and-forth behavior at intermediate light intensities. The existence of this threshold also enables to explain why the time scale of negative phototaxis is longer than that of positive phototaxis. Negative phototaxis occurs when the amount of active species $\sbound$ overcomes a threshold $s_T$; this takes time, during which the alga exhibits  positive phototaxis. Only when $\sbound>s_T$ do the algae show negative phototaxis, and accumulate at one boundary of the well. The small initial positive phototactic response predicted by the model is in agreement with our experimental data \reviewerone{and with recent data reported in~\cite{catalan2023light}.} \reviewerthree{The model is also qualitatively compatible with the history-dependent trajectories of \chlamy\ reported in~\cite{arrieta2017phototaxis}. There, the authors report a diaphototactic behavior of \chlamy\ which cannot be reproduced by a simple switch from positive to negative phototaxis at a threshold light intensity. Simulating trajectories of microalgae whose phototactic behavior depends on their history of illumination enables to recover a  looping trajectory around a gaussian light source, qualitatively similar to that reported in~\cite{arrieta2017phototaxis}, see Supp. Fig.~14. Further simulations and experiments are needed to obtain quantitative results and confirm the adequation of our model with the trajectories in~\cite{arrieta2017phototaxis}.}

The model also predicts that it is possible to switch the phototactic behavior of algae from positive phototaxis to back-and-forth to negative phototaxis, using three consecutive identical stimuli, by adjusting the waiting times between the stimuli and choosing carefully the stimulus intensity, see Fig.~\ref{fig:osci model/model_3_phototaxis_paused}d \reviewerone{and Supp. Fig.~11d}. We were not able to find the experimental conditions corresponding to this theoretical prediction,  but did experimentally observe the switch from back-and-forth to negative phototaxis, and from positive phototaxis to back-and-forth in a couple experiments, see Supp. Fig.~13.

While the model successfully reproduces the phototactic behavior and the time scales at play, it fails to reproduce two experimental observations. First, in a couple experiments, multiple consecutive back-and-forth of the center of mass were observed at intermediate intensities, see Supp. Fig.~15. Second, in experiments with consecutive stimuli at intermediate light intensities, algae in the third experiment (after a waiting time of 30 minutes) display positive phototaxis instead of the expected back-and-forth behavior observed during the first experiment, see Fig.~\ref{fig:time}b and Supp. Fig.~13. Both behaviors can be obtained in the model by supposing that the threshold $s_T$ increases with the time spent under a light stimulus, and so depends on the history of the microorganism. Such an increasing threshold would also enable to recover  the traditional adaptative behavior at longer time scales~\cite{mayer1968chlamydomonas}. This implies having a third equation describing the time-evolution of $s_T$, thus introducing another time constant into the problem. \reviewertwo{Such a model would still be an over-simplification of the biological reality, where multiple pathways are activated in parallel, and different types of photoreceptors are activated in response to light and participate in phototaxis; for example, the existence of DYBLUP/MOT7, a dynein-associated photoreceptor was recently exhibited. DYBLUP prevents phototactic adaptation of algae subjected to intense blue light~\cite{kutomi2021dynein}, an adaptation that otherwise occurs in mutants lacking MOT7 with a time scale of $\approx 30$~minutes. This type of detailed modelling is out of the scope of this article.}

 In the lab, the memory effects reported here imply that it is not possible to repeat the same phototaxis experiments on a given batch of cells and expect similar outcomes. Such an effect was already known in the case of long-term adaptation~\cite{mayer1968chlamydomonas}. The data reported here show that even in the case of short-term experiments, the behavior is not necessarily reproducible. To study phototaxis quantitatively, experimental repeats should be carried out with different batches of naive cells. Having identified this short time-scale, our experiments pave the way for the study of phototaxis in the presence of time-varying light stimuli, much as the algae are subject to in their natural environment. There, light is constantly fluctuating. The short-term memory could then be a beneficial way to integrate a light signal randomly interrupted by the shadow of other motile organisms or objects.

\section*{Acknowledgments} 

We thank David Gonzalez-Rodriguez for insightful comments on an early version of the model, Maxime Simon for preliminary experiments, Salome Guttierez-Ramos for help with microfabrication, Caroline Frot for precious technical support and Julien Bouvard for help with calibration measurements. This work was supported by ``Investissements d’Avenir'' LabEx PALM (ANR-10-LABX-0039-PALM) and by PEPS ``M\'{e}canique du futur''.

\bibliographystyle{apsrev4-2}
\bibliography{biblio}% Produces the bibliography via BibTeX.

\clearpage
\newpage
\appendix
\setcounter{figure}{0} 
\section*{Supplementary Material}
\renewcommand{\figurename}{Supp. Fig.}

\section{Material and Methods}
\subsection{Concentration of the algal solution}

The algal culture was taken 4 hours after the beginning of the ``day", and underwent three centrifugation steps, leading to a solution concentrated in motile algae, and enabling to get rid of low-motility algae, dead algae, and cellular debris. First, 45 mL of the liquid culture were centrifuged at 1057g for 10 minutes. Then, 39 mL of the supernatant were removed to obtain a concentrated pellet of cells. The bottom 6 mL were homogenized and then centrifuged at 73g for 2 minutes. The supernatant, containing the motile cells, was kept and centrifuged again at 285g for 5 minutes to obtain a final solution highly concentrated in motile algae. This solution was then diluted at the desired concentration for the experiments. Before experiments, algae were left to rest in the dark for 60 minutes, allowing the cells that had deflagellated during the centrifugation process to regrow their flagella \cite{rosenbaum1969flagellar, lefebvre1978flagellar}. It is known that the phototactic response of microalgae may be regulated by its inner biological circadian clock through the day \cite{bruce1970biological}. To ensure the reproducibility of our experiments, experiments systematically started 6 hours after the beginning of the ``day".

\subsection{Projected area fraction $\phi$}

\begin{figure}[htb!]
  \centering
  \includegraphics[width=0.8\linewidth]{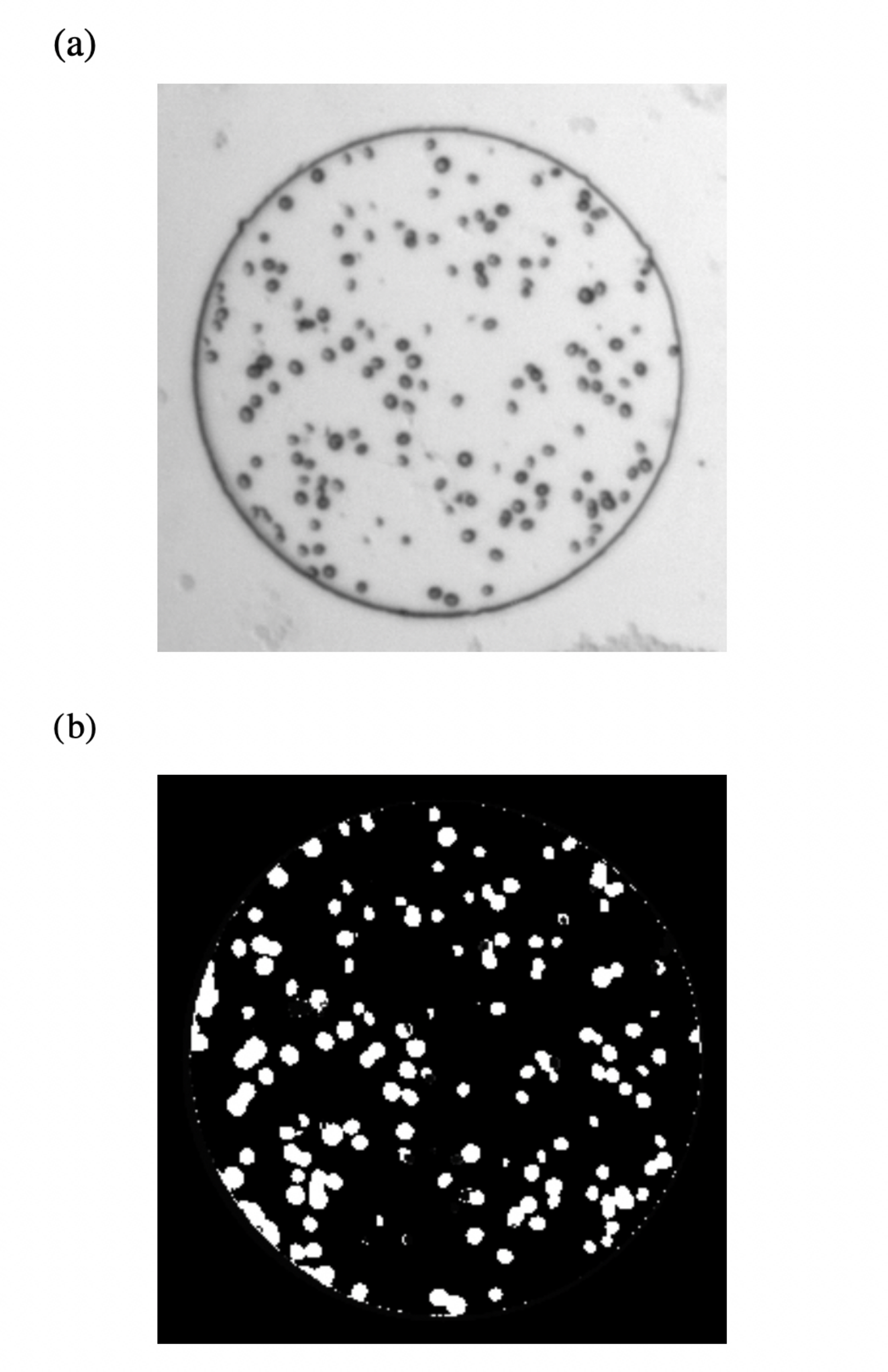}%
    \centering
    \caption{Binarization of the experimental images using a threshold on pixel intensity. (a) An experimental image of a well with algae in it. The algae are darker than the background. (b) Thresholding the experimental image leads to a binarized image where algae appear in white in a dark background.}
    \label{fig:imageThresholding}
\end{figure}

\begin{figure*}[htb!]
  \centering
  \includegraphics[width=1\linewidth]{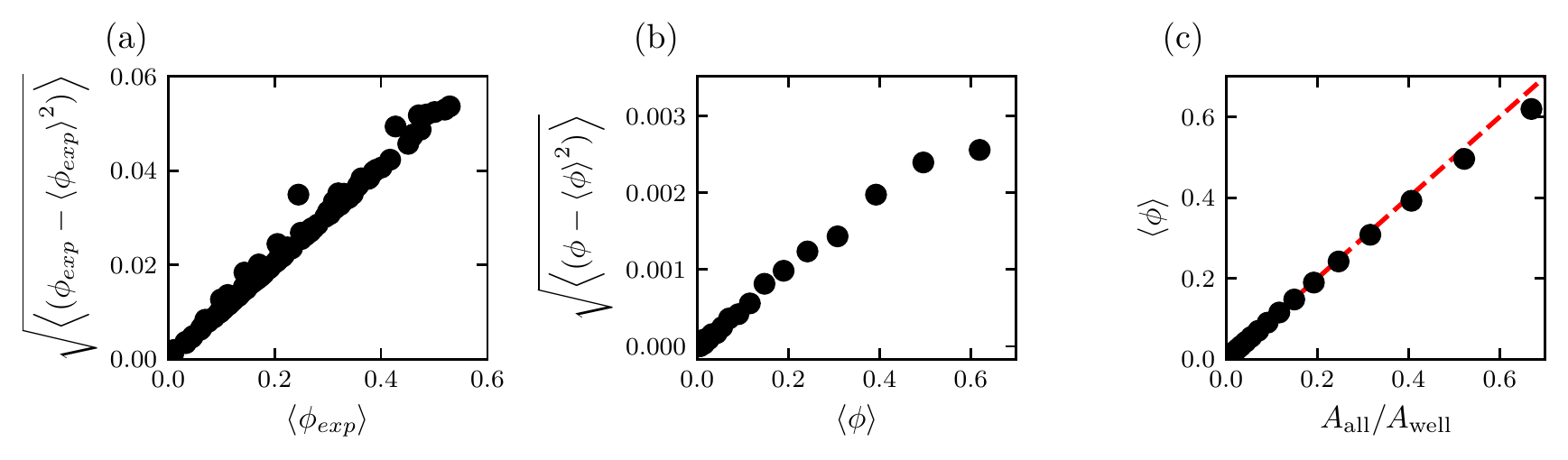}%
    \centering
    \caption{ (a) The error in the concentration measurement of each experiment is quantified by calculating the standard error in the projected area of algae in the first hundred images before the light is turned on. The relative average standard error is $10\%$ of the concentration $\phi$. Points: experiments. The dashed red line has a slope of 0.1. (b) Checking for overlaps: projected area fraction $\phi \equiv A_p/A_{\rm well}$ of $N$ confined spheres which can overlap in the $z$ direction, as a function of the  projected area $A_{\rm all}/A_{\rm well}\equiv N\pi R^2/A_{\rm well}$ of $N$ spheres of radius $R$. The red dashed line has slope 1, showing that overlaps can essentially be neglected for $\phi \leq 0.7$. (c) Standard deviation of the projected area fraction $\phi \equiv A_p/A_{\rm well}$ as a function of the average  projected area fraction $\left<\phi\right>$. The average and standard deviation are calculated over 100 simulations. }
    \label{fig:asupp_error_concentration_measurement}
\end{figure*}

The concentration in algae in each well was determined by calculating, in each well, the fraction of area occupied by the algae. At the beginning of an experiment, the algae were not stimulated by any blue light, and swam randomly in their wells. We used Otsu thresholding to binarize the images, see Supp. Fig.~\ref{fig:imageThresholding} and obtain, for each well, the total area occupied by the algae $A_p$. In each well, this area was renormalized by the well area $A_{\rm well}$. We then defined the projected area fraction $\phi \equiv A_p/A_{\rm well}$, which was used as a proxy for the concentration in algae. This was repeated for the first 100 images of each experiments, and used to calculate the mean value and the standard deviation of $\phi$. We find that the relative error is of the order of  10\%, see Supp. Fig.~\ref{fig:asupp_error_concentration_measurement}a.

The uncertainty of 10\% on $\phi$ in our experiments results from the imperfect binarization of images, and not from the fact that algae overlap in $z$. Indeed, at the values of $\phi \leq 0.5$ used in the experiments, there is almost no overlap. This can also be checked by simulating $N$ solid spheres with a radius $R=8\; \rm \mu m$ placed randomly in a cylindrical well of height $32\; \rm \mu m$, and calculating their projected area. The obtained projected area is equal to the projected area of $N$ spheres as long as $\phi\leq 0.7$, see Supp. Fig.~\ref{fig:asupp_error_concentration_measurement}b. The theoretical error on $\phi$, determined by calculate the standard deviation of $\phi$ in 100 identical simulations, is of the order of 0.5\%, much smaller than the experimental error, see Supp. Fig.~\ref{fig:asupp_error_concentration_measurement}c.

\subsection{Determining the flux of photons seen by the microalgae}

To determine the flux of photons reaching the microalgae, we proceeded in two steps.

For all experiments, we measured the flux of photons from the blue LED reaching the camera sensor. To do so, we first determined the camera offset value by blocking off all the light to the camera and taking a 16-bit image. The spatial average intensity in grey value of all the pixels in the image was the camera's offset. Then, a 16-bit setup image of the sample was taken at the current experimental condition, with the light of the microscope turned off and the blue LED light on. The camera offset value was then subtracted from each pixel in the setup image. The grey values were first converted to number of electrons by dividing each pixel in the image by the conversion gain of the camera. The electrons were converted to photons by dividing the number of electrons from each pixel by the quantum efficiency (QE) of the sensor at $\lambda = 470\; \rm nm$.

\begin{figure}[htb!]
  \centering
  \includegraphics[width=\linewidth]{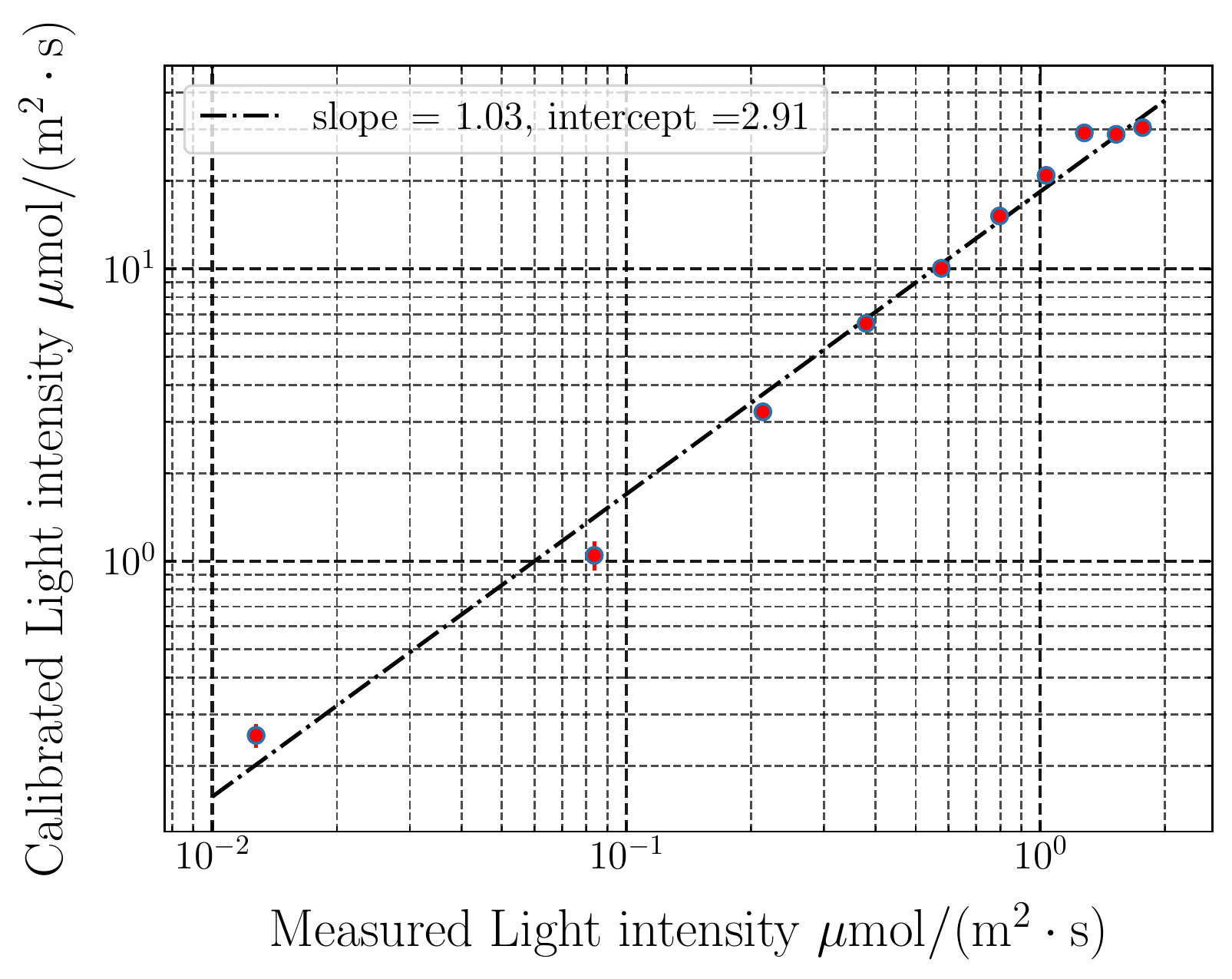}%
    \centering
    \caption{Calibration of the light intensity measurement. The light scattered by the PDMS is measured at the level of the camera sensor using the gray values of the recorded images (x-axis). The light intensity at the level of the microwells is measured using a light sensor (y-axis). Both values are proportional, with a coefficient of proportionality $\approx 20$. }
    \label{fig:calibrationCurve}
\end{figure}

Note that the amount of light reaching the camera  corresponds to the light scattered by the PDMS. To relate it to the light stimulus experienced by the algae, we measured once the light intensity at the level of the PDMS chip using a light sensor (Adafruit TSL2591), connected to an Arduino. Relating this light intensity to the  intensity recorded by the camera provides a calibration curve, enabling to determine the flux of photons reaching the algae. This calibration curve shows that the flux of photons reaching the algae is 20 times higher than the one scattered towards the camera sensor, see Supp. Fig.~\ref{fig:calibrationCurve}

\section{Experimental results}

\subsection{Fraction of algae not responding to light}

\begin{figure}[htb!]
  \centering
  \includegraphics[width=1\linewidth]{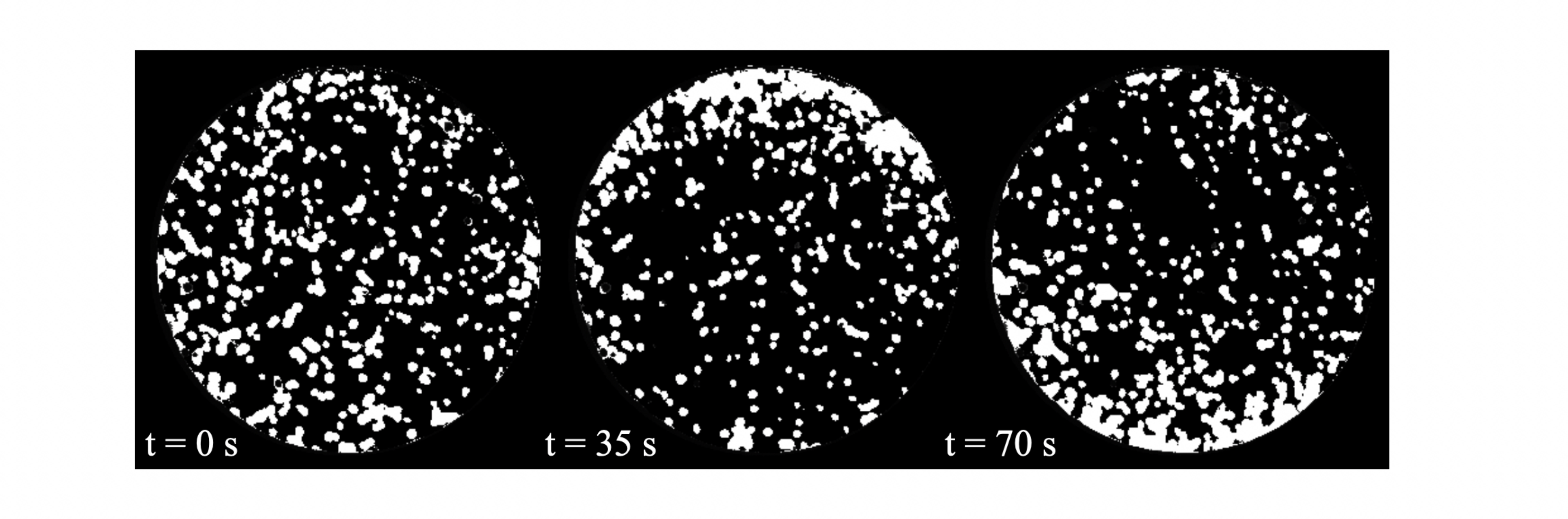}%
    \centering
    \caption{Time-lapse of binarized images of \chlamy\ enclosed in a well. The blue light stimulus is turned on at $t = \rm 30s$. The stimulus comes from the upper side of the well. Not all algae react to the stimulus.}
    \label{fig:supp_thresholded_osci}
\end{figure}

Not all algae respond to the light stimuli, see the time-lapse in Supp. Fig.~\ref{fig:supp_thresholded_osci}. To determine the fraction $f$ of algae  responding to light, we compare the number of algae that have accumulated at the edge of the well at the end of the experiment with the total number of algae.

\begin{figure*}[htb!]
  \centering
  \includegraphics[width=1\linewidth]{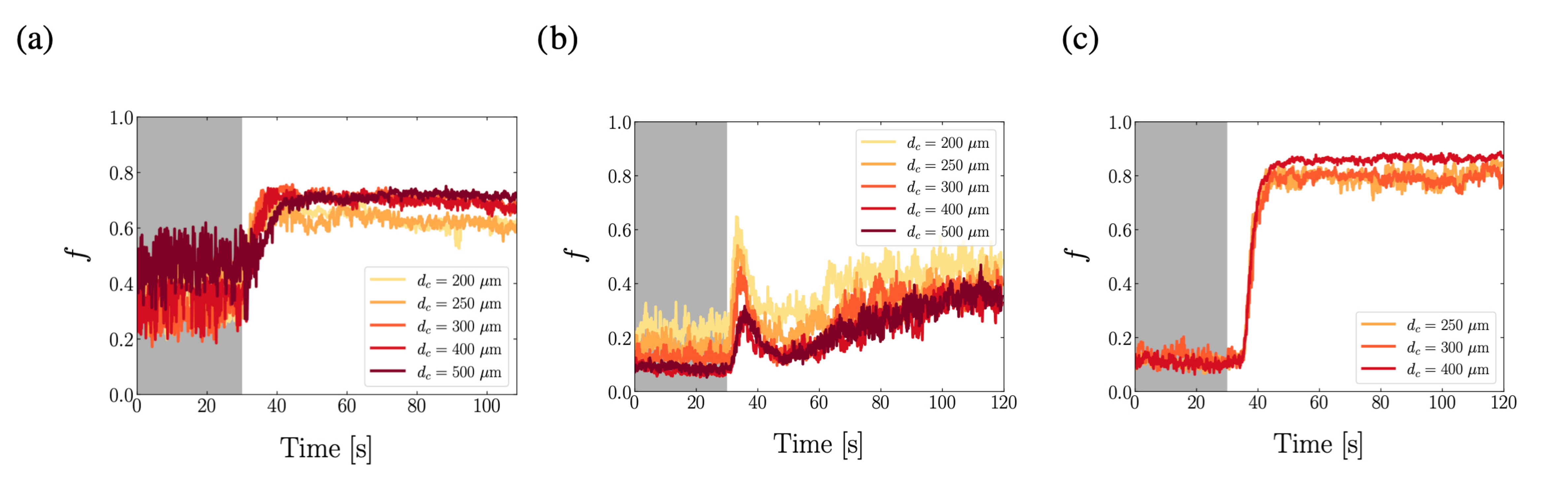}%
    \centering
    \caption{The light stimulus is turned on at $t=30\; \rm s$ and the fraction $f$ of algae accumulated at the wall, and therefore that react, is tracked over time. (a) $I = 0.38 \; \mmol $, the algae show positive phototaxis. 75\% of the algae react to the light stimulus and accumulate at the wall. (b) $I = 2.4 \; \mmol$, the algae show first positive phototaxis, then negative phototaxis. At most, 65\% of the algae react to the light stimulus and accumulate at the wall during the transient positive regime and 50\% during negative phototaxis. (c) $I = 28 \; \mmol$, the algae show negative phototaxis. 89\% of the algae react to the light stimulus and accumulate at the wall. Each curve is the fraction of accumulated algae averaged over 5 experiments. These are the same experiments as in Fig. 3 of the main article.}
    \label{fig:accumulation_dynamics_diameter}
\end{figure*}

The fraction $f$ of responding algae is shown in Supp. Fig.~\ref{fig:accumulation_dynamics_diameter} for the same experiments as in Fig. 3 of the main article. Note that $f$ is defined with respect to the largest connected component of the image, and so has no meaning before $t\leq 30 \; \rm s$, when the algae are not accumulated.

\begin{figure*}[ht!]
  \centering
  \includegraphics[width=1\linewidth]{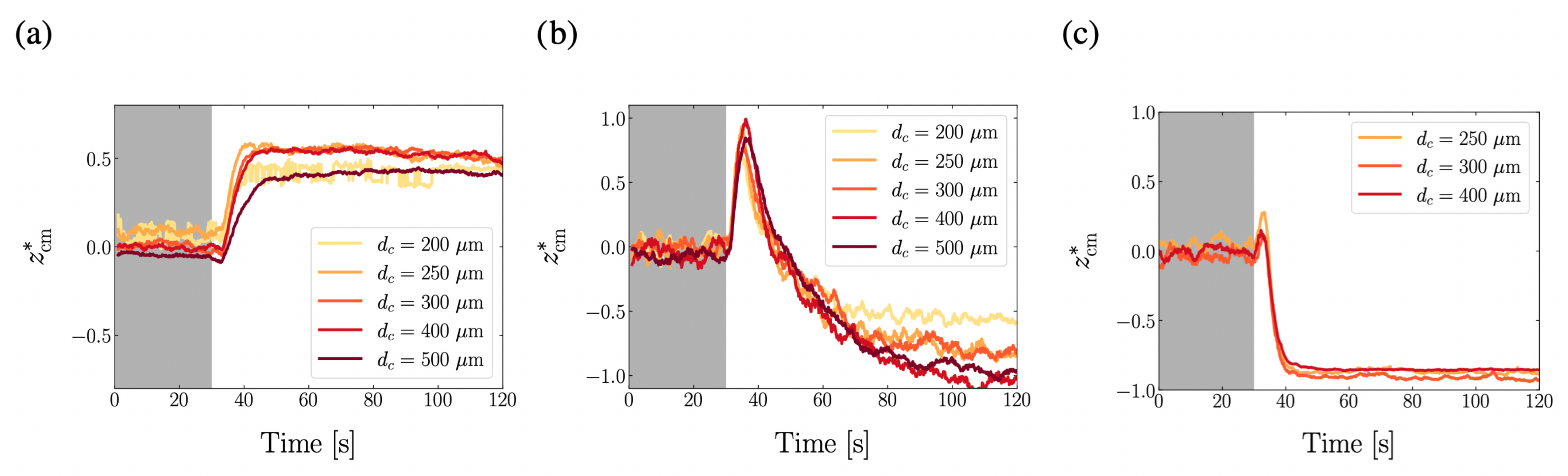}%
    \centering
    \caption{The barycenter $\zcm$ is renormalized by the reacting fraction of algae. The light stimulus is turned on at $t=30\; \rm s$ and the corrected position of the center of mass $z_{cm}^{\ast}$ is tracked over time. (a) When exposed to an intensity $I = 0.38 \; \mmol $, the algae show positive phototaxis. The average projected area fraction of algae in 30 wells is $\phi_{avg}=0.36$. (b) At intermediate intensities $I = 2.4 \; \mmol$, the algae show first positive phototaxis, then negative phototaxis. The average projected area fraction of algae in 30 wells is $\phi_{avg}=0.28$. (c) At a high intensity $I = 28 \; \mmol$, the algae show negative phototaxis. Each curve is an average over 4 to 7 experiments. The average projected area fraction of algae in 17 wells is $\phi_{avg}=0.11$.}
    \label{fig:rescaled}
\end{figure*}

The fraction of responding algae $f$ can then be used to renormalize the value of the center of mass, $\zcm^{\star} = f^{-1}\cdot \zcm $. Results of the renormalization are shown in Supp. Fig.~\ref{fig:rescaled}. Values of $\zcm^{\star}$ are much closer to $\pm 1$ than values of $\zcm$, showing that the main cause of the center of mass not going to $\pm 1$ are the immobile algae.

\subsection{High concentrations of algae}

\begin{figure*}[htb!]
  \centering
  \includegraphics[width=1\linewidth]{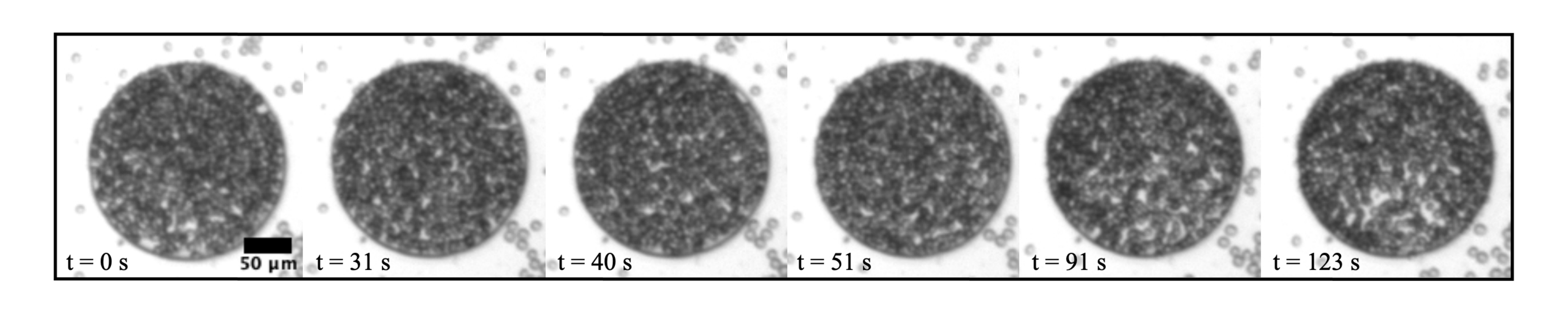}%
    \centering
    \caption{\chlamy\ enclosed in a well. The blue light stimulus is turned on at $t=30\; \rm s$. The stimulus comes from the upper side of the wells. The algal population fills the entire well, preventing the algae from migrating towards or away from the light.}
    \label{fig:high_concentration_well}
\end{figure*}

At too high concentrations, the algae fill the entire well, preventing the algae from swimming towards or away from the light, see Supp. Fig.~\ref{fig:high_concentration_well}.

\subsection{Sticking algae}

\begin{figure*}[ht!]
  \centering
  \includegraphics[width=1\linewidth]{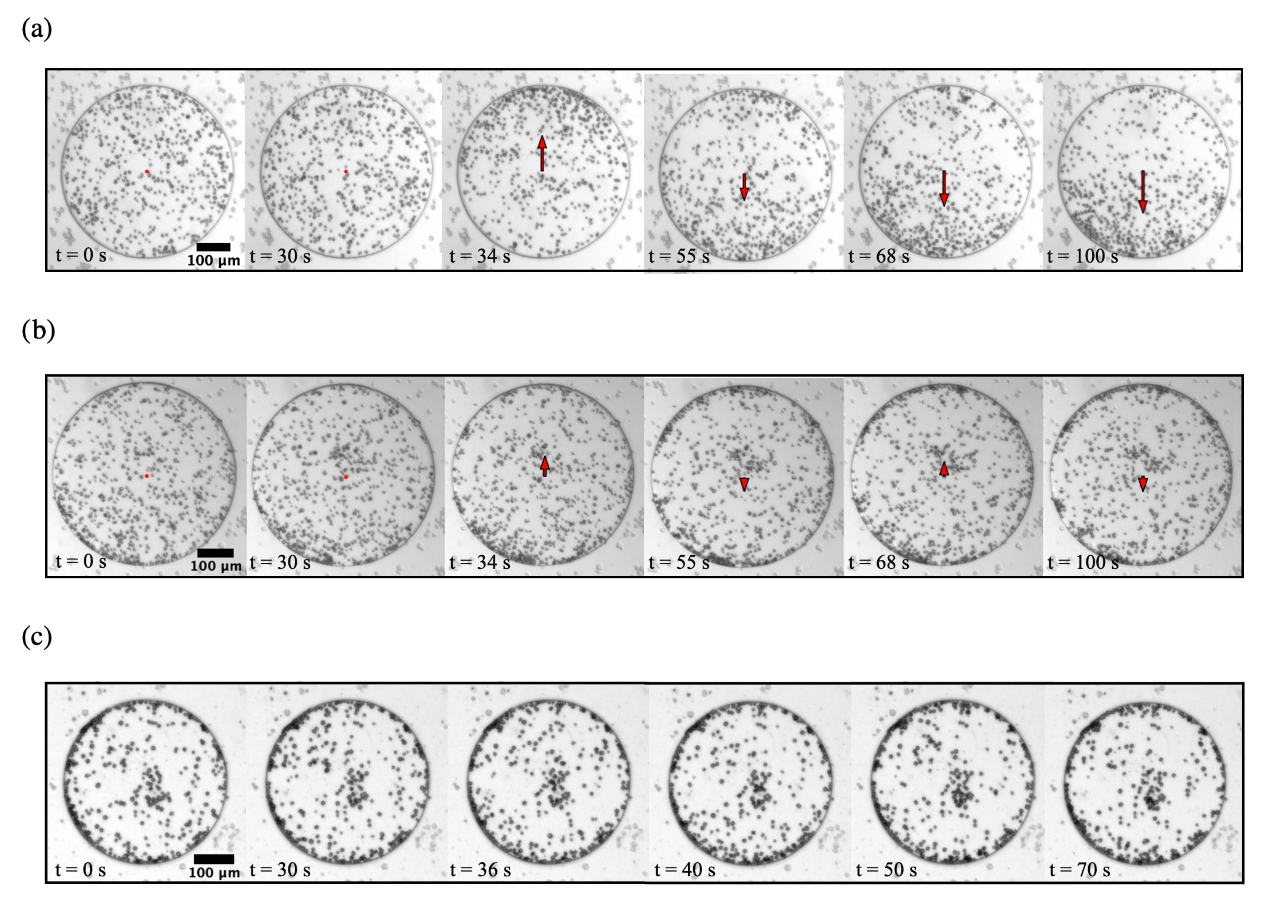}%
    \centering
    \caption{\chlamy\ enclosed in a well. The blue light stimulus is turned on at $t=30\; \rm s$. The stimulus comes from the upper side of the wells. (a) In the first experiment performed at intermediate intensity $I = 1.78 \; \mmol$, the algae show first positive phototaxis, then negative phototaxis. The algae are motile. (b) At the beginning of the fourth experiment performed at intermediate intensity $I = 1.78 \; \mmol$ many algae remain accumulated at the walls from earlier experiments. We can see aggregates forming at the center of the well and at the walls during the course of the experiment. (c) The sixth experiment of a series performed at intensity $I = 6.6 \; \mmol$ is displayed. The algae have formed aggregates at the wall and at the center and do not react to the light stimulus anymore.}
    \label{fig:stuck_algae}
\end{figure*}

After repeated stimuli, the algae can stick to the glass, see Supp. Fig.~\ref{fig:stuck_algae}. This has already been reported, see~\cite{catalan2023light,kreis2018adhesion}. 

\subsection{Influence of the time between experiments on the change in phototactic behavior}

\begin{figure*}[htb!]
  \centering
  \includegraphics[width=1\linewidth]{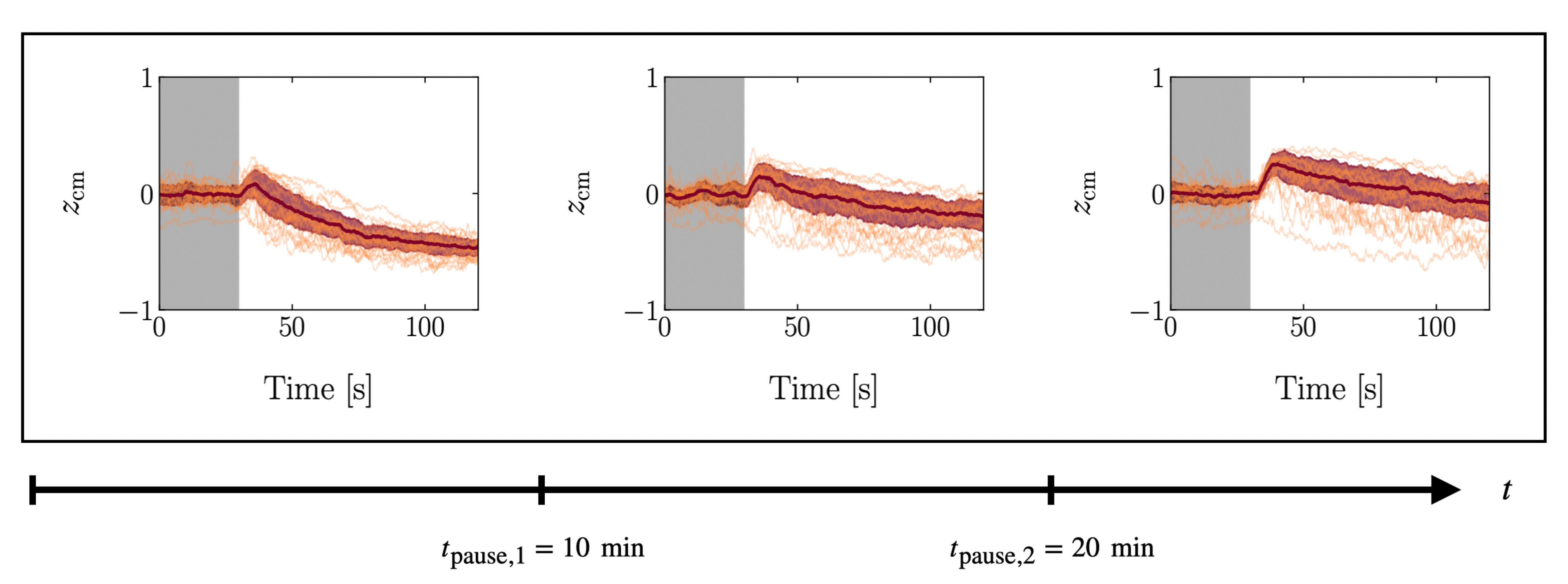}%
    \centering
    \caption{Memory effects on the phototactic response of algae. The light stimulus is turned on at $t=30\; \rm s$ and the position of the center of mass $\zcm$ is tracked over time. Three consecutive experiments are performed with varying rest times, where the  light is turned off in between. There is a 10 min pause between Exp. 1 and Exp. 2, then a 20 min pause between Exp. 2 and Exp. 3. At intermediate intensities $I = 1.3\; \mmol$, the algae show the same phototactic behaviors in each experiment. The center of mass is averaged over 24 experiments.}
    \label{fig:effect_of_history_5-10min_breaks}
\end{figure*}

We stimulated populations of algae with a stimulus eliciting as a first response a back-and-forth behavior. Repeating the stimulus at a 10 or 20 minutes interval did not lead to a change in the phototactic response, see Supp. Fig.~\ref{fig:effect_of_history_5-10min_breaks}.

\begin{figure*}[ht!]
  \centering
  \includegraphics[width=1\linewidth]{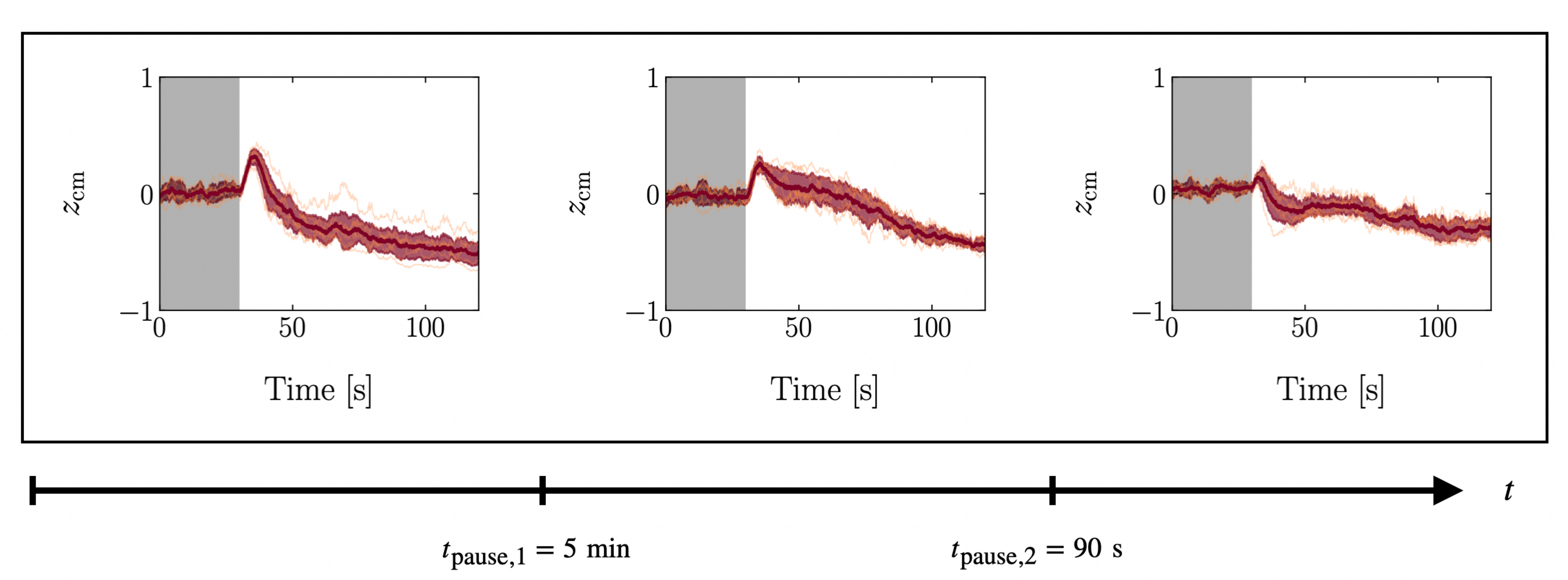}%
    \centering
    \caption{Memory effects on the phototactic response of algae. The light stimulus is turned on at $t=30\; \rm s$ and the position of the center of mass $\zcm$ is tracked over time. Three consecutive experiments are performed with varying rest times, where the  light is turned off in between. There is a 5 min pause between Exp. 1 and Exp. 2, then a 90 s pause between Exp. 2 and Exp. 3. At intermediate intensities $I = 2.4\; \mmol$, the algae the same phototactic behavior in the first two experiments, there is a back-and-forth motion. In the last experiment, the algae switch to negative phototaxis after a short 90 s break. The center of mass is averaged over 6 experiments.}
    \label{fig:effect_of_history_5-2min_breaks}
\end{figure*}

Repeating the stimulus after a 5 minute interval also does not lead to a change in the phototactic sign, see the first two graphs in Supp. Fig.~\ref{fig:effect_of_history_5-2min_breaks}. Repeating the stimulus after 90 seconds leads to a change from back-and-forth to negative photaxis, see last graph in Supp. Fig.~\ref{fig:effect_of_history_5-2min_breaks}.

%%%%%%%%%%%%%%

\section{Simplified phototactic model}
\subsection{Determining the parameters of the simplified phototactic model}

The dynamics of the inner biochemical species $S$ in our simplified phototaxis model evolve according to:

\begin{equation}
    \frac{d\sbound}{dt} = \gamma I_0 (\stot-\sbound) - \tau^{-1}\sbound,
    \label{eq:sEq}
\end{equation}

where $\gamma$ is the reaction rate at which the inactive species of concentration $\sfree$ is converted into the active species of concentration $\sbound$. The total concentration is conserved and is called $\stot = \sfree + \sbound$.  The transition from inactive to active state depends on the light intensity $I_0$, while the reverse transition occurs at a constant rate $\tau^{-1}$.

The solution to this equation writes
\begin{equation}
    \sbound(t) = \frac{\gamma I_0 \tau}{\gamma I_0 \tau + 1}\stot\left(1-{\rm exp}\left\{-\left[\frac{\gamma I_0 \tau + 1}{\tau}\right]t \right\}\right).
\end{equation}

We use $\stot = 1$ for simplicity. To obtain time scales close to the experimental time scales, we choose $\tau = 5 \; \rm min$.

Then, we assume the position $z$ of the alga evolves according to:
\begin{equation}
    \frac{dz}{dt} = -{\rm sign}(\sbound - s_T)v_0,
    \label{eq:xEq}
\end{equation}
where $s_T$ is the threshold concentration at which cells transition from positive to negative phototaxis, and $v_0$ is the characteristic speed of the alga. We know experimentally that $v_0 \approx 100 \; \rm \mu m.s^{-1}$. 

To obtain the phase diagram of the phototactic behavior as a function of $\gamma I_0$ and $s_T$ shown in the main text of the article, we defined a positive phototactic behavior when $\sbound(t) < s_T$ for $0\leq t \leq 90\; \rm s$, and negative phototactic behavior when $\sbound(t)$ crosses the threshold $ s_T$ at one point $t$ such that $0\leq t\leq = 10 \; \rm s$. In between, $\sbound(t)$ crosses the threshold $s_T$ at a time $10\; rm s\leq t\leq 90; \rm s$, and this defines a back-and-forth behavior.

Two parameters now need to be determined: $\gamma$ and $s_T$. To estimate $\gamma$, we use the fact  that, at $I_0 = 0.02\; \mmol$, the algae stop responding. We assume this corresponds to less than one molecule of activated $S$ per second~\cite{ramamonjy2022nonlinear}, leading to $\gamma = 2\times 10^{-3} \; \rm m^2 \cdot \mu mol^{-1}$. Then, we also know that at $I_0 = 2 \; \mmol$, the algae exhibit a back-and-forth behavior. This determines $s_T = 0.1$, according to the phase diagram in the main text.

\subsection{Concentration of $\sbound$ for different stimuli}

The sign of phototaxis is given in the model by comparing the concentration $\sbound$ of activated chemical, to a threshold value $s_T$. The evolution of $\sbound$ with time for repeated stimuli of different intensities is shown in Supp. Fig.~\ref{fig:osci model/model_3_phototaxis_paused}. 

\begin{figure*}[ht!]
  \centering
  \includegraphics[width=1\linewidth]{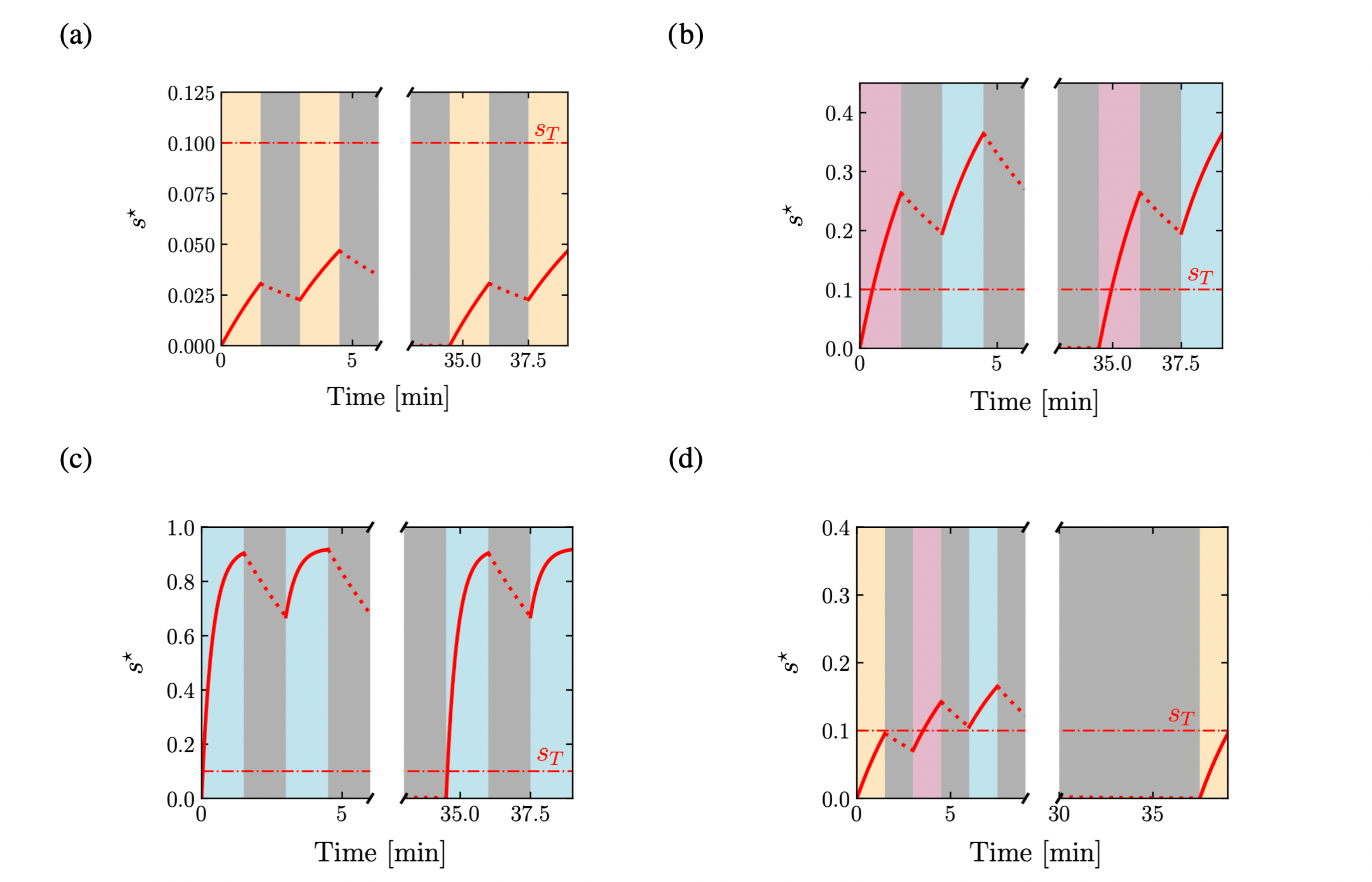}%
    \centering
    \caption{Simulation results for four consecutive experiments performed at a constant light intensity $I_0$, with different rest times in between. The concentration of active molecules $s^{\ast}$ is tracked over time. Each experiment lasts for $90$ s. Rest times are shown in shaded gray areas. ($t_{\rm{pause,1}}=90$ s, $t_{\rm{pause,2}}=30$ min and $t_{\rm{pause,3}}=90$ s for (a), (b) and (c)). (a) When exposed to a low light intensity $I_0 =0.2\; \mmol$, the alga always displays positive phototaxis. (b) At intermediate intensities $I_0 = 2\; \mmol$, the alga varies its phototactic behavior between each experiment. In the first experiment, the alga shows a back-and-forth motion. In the second, the alga shows only negative phototaxis. In the third experiment, the alga displays a back-and-forth motion again. Finally, in the last experiment, the behavior is negative phototaxis. (c) At a high intensity $I_0=20 \; \mmol$, the alga always shows negative phototaxis. (d) At low to intermediate intensities $I_0 = 0.65\; \mmol$ and with adjusted pause times ($t_{\rm{pause,1}}=90$ s, $t_{\rm{pause,2}}=90$ s and $t_{\rm{pause,3}}=30$ min) it is possible to successively go through positive phototaxis in the first experiment, back-and-forth motion in the second, negative phototaxis in the third and then revert back to positive phototaxis in the last experiment. Other simulation parameters: $s_{T} = 0.1$, $\stot = 1$, $v_0=100 \; \mu \rm m \cdot \rm s^{-1}$, $\tau = 300\; \rm s$ and $\gamma = 0.002\; \rm m^2\cdot  \mu mol^{-1}$.}
    \label{fig:osci model/model_3_phototaxis_paused}
\end{figure*}

\subsection{Response of the phototactic model to consecutive stimuli}

\begin{figure}[htb!]
  \centering
  \includegraphics[width=1\linewidth]{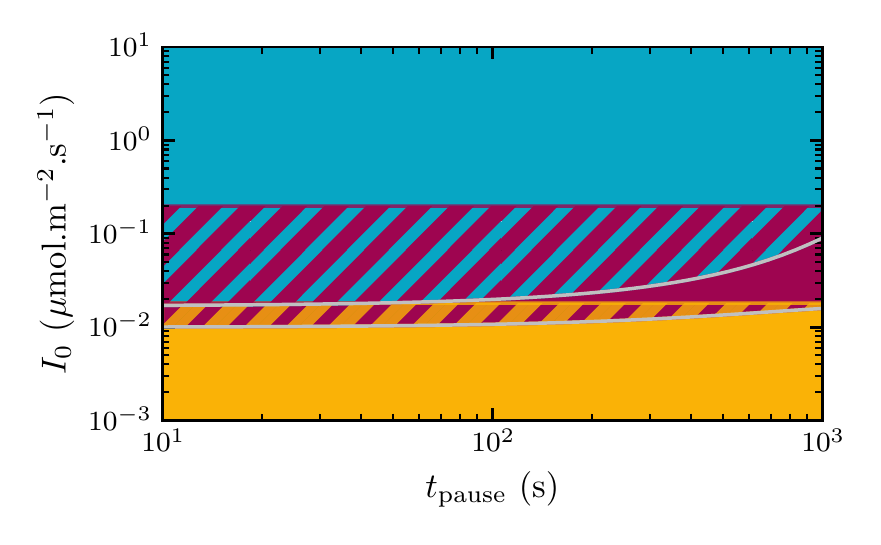}%
    \centering
    \caption{Phase diagram of the phototactic behavior after two consecutive stimuli at the same intensity. Yellow: positive phototaxis. Red: Back-and-forth behavior. Blue: Negative phototaxis. Hatched regions indicate where the behavior changes between the first and second stimulus. Yellow and red hatch: positive phototaxis during the first stimulus, back-and-forth during the second stimulus. Blue and red hatch: back-and-forth during the first stimulus, negative phototaxis during the second stimulus. Simulation parameters: $\gamma = 2\times 10^{-3} \;  \rm m^2\cdot  \mu mol^{-1}$, $s_T = 0.1$, $\tau = 300\; \rm s$.}
    \label{fig:phaseDiagramRestimulation}
\end{figure}

We simulated the application of two consecutive, identical stimuli of intensity $I_0$ and duration 90 seconds, spaced by a time $t_{\rm pause}$. We can then draw the phase diagram showing when the algae change behavior between the two stimuli. This phase diagram is shown in Supp. Fig.~\ref{fig:phaseDiagramRestimulation}. In the diagram, regions filled with a unique color show when the behavior does not change between the two stimuli. Yellow: positive phototaxis. Red: back-and-forth behavior. Blue: negative phototaxis. Regions filled with hatched lines indicate a change in behavior between the two stimuli. Yellow and red hatches: the algae exhibit positive phototaxis in the first stimulus and back-and-forth in the second stimulus. Red and blue hatches: the algae exhibit back-and-forth motion in the first stimulus and negative phototaxis in the second stimulus.

\subsection{Multiple changes in phototactic behavior}

\begin{figure*}[htb!]
  \centering
  \includegraphics[width=1\linewidth]{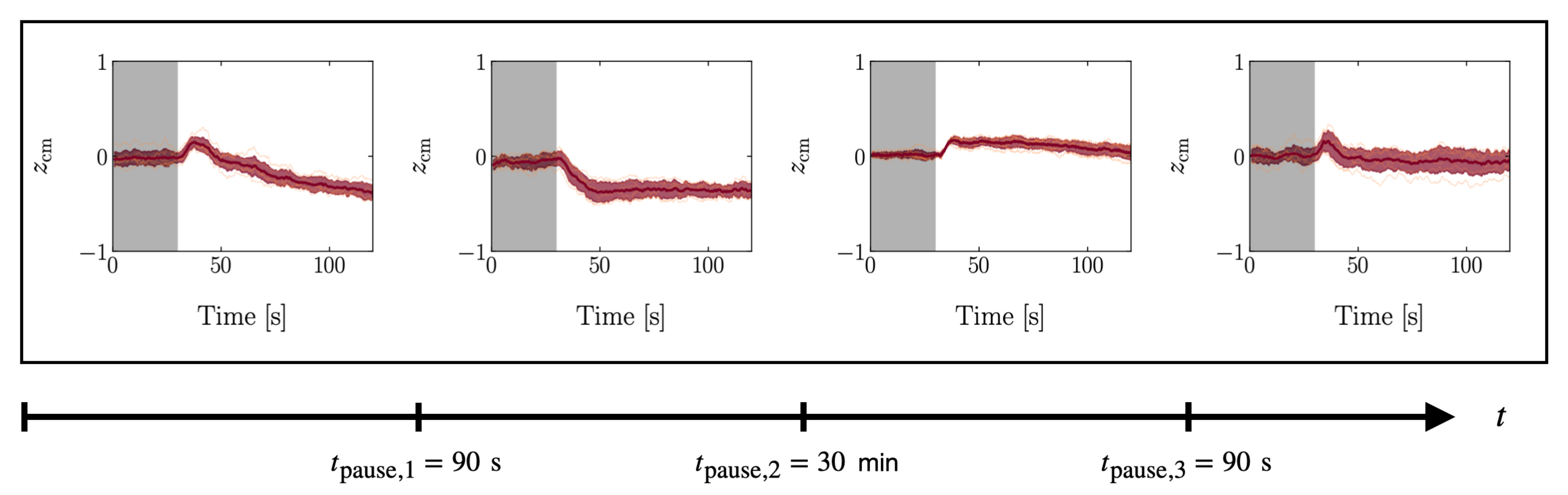}%
    \centering
    \caption{ Memory effects on the phototactic response of algae. After a pause of 90 seconds, it is possible to switch from back-and-forth behavior to negative phototaxis (Exp.~1 and 2), or from  positive phototaxis to back-and-forth (Exp.~3 and 4). The light stimulus is turned on at $t=30\; \rm s$ and the position of the center of mass $\zcm$ is tracked over time. Four consecutive experiments are performed with varying rest times, where the  light is turned off in between. There is a 90 s pause between Exp. 1 and Exp. 2, then a 30 min pause between Exp. 2 and Exp. 3 and finally a 90 s pause between Exp. 3 and Exp. 4. At intermediate intensities $I = 0.46\; \mmol$, the algae show different phototactic behaviors in each experiment. The behavior switches from positive phototaxis to back-and-forth behavior between the third and fourth experiments. The center of mass is averaged over 6 experiments.}
    \label{fig:effect_of_history_supp}
\end{figure*}

It is possible to switch from back-and-forth to negative phototaxis, and from positive phototaxis to back-and-forth in successive experiments separated by a short 90 s break, see Fig.~\ref{fig:effect_of_history_supp}.

\subsection{Incorporating the inner biochemistry into the model of Arrieta et al.~\cite{arrieta2017phototaxis}}

\begin{figure}[htb!]
  \centering
  \includegraphics[width=1\linewidth]{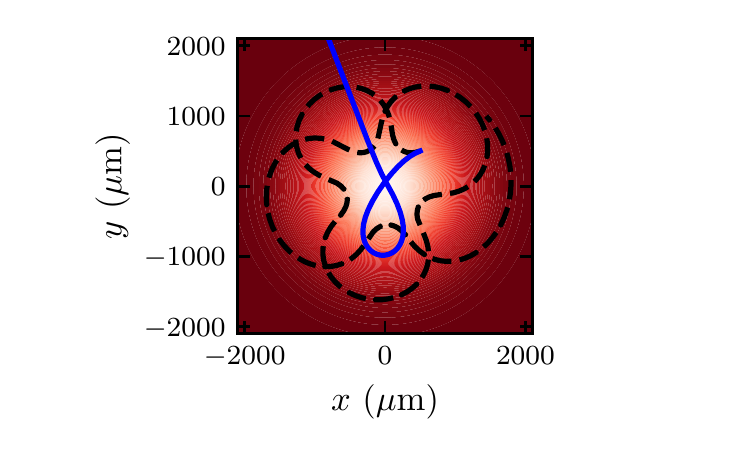}%
    \centering
    \caption{Incorporating memory in the model of Arrieta et al.~\cite{arrieta2017phototaxis}. Black dotted line: simulated trajectory of a cell that exhibits positive phototaxis at light intensities $I < I_c$, and negative phototaxis otherwise. The shape of the trajectory is not the shape observed in experiments. Blue line: simulated trajectory of a cell whose phototactic behavior depends on the concentration $\sbound$ of an inner biochemical species, with a characteristic deactivation time $\tau = 300\; \rm s$. The cell makes a loop and then escapes.  }
    \label{fig:arrieta}
\end{figure}

It is possible to incorporate the dynamics of $\sbound$ into another model of phototaxis, described in Arrieta et al.~\cite{arrieta2017phototaxis}. There, the authors report that \chlamy\ describe loops around gaussian light sources, before escaping. They show that this behavior cannot be reproduced by a simple phototaxis model where the sign of phototaxis changes at a threshold intensity $I_c$. Arrieta et al. assume the position of a cell $\bm{{x}}(t)$ and its direction $\bm{p}(t)$ evolve according to:
\begin{equation}
    \bm{\dot{x}}(t) = v_s\bm{p}(t) \qquad \mbox{ and } \qquad \bm{\dot{p}}(t) = \bm{\omega}\times \bm{p}(t) ,
\end{equation}
where $v_s$ is the speed of a cell and $\bm{\omega}$ is its angular speed. The angular speed is supposed to be proportional to the local gradient in light intensity $\nabla I$: $\bm{\omega} = \alpha \bm{p}(t) \times \nabla I$, where $\alpha$ is the phototactic parameter. In a simple assumption, $\alpha = 1$ (resp. $-1$) when the local light intensity is below (resp. above) a threshold $I_c$. This leads to the trochoid-like trajectory shown as a dotted black line in Supp. Fig.~\ref{fig:arrieta}. We now incorporate our model of the dynamics of $\sbound$ into the phototactic parameter, and assume $\alpha = 1$ (resp. $-1$) when $\sbound \leq s_T$ (resp. $\sbound > s_T$). For a given set of parameters, this leads to the blue trajectory in  Supp. Fig.~\ref{fig:arrieta}: the algae loops around the light and escapes. The escape is due to the memory: due to a too long exposure to intense light, \chlamy\ becomes negatively phototactic during a time $\approx \tau$. After this time, it has swam far away from the source, and does not feel the gradient of light anymore so does not come back towards the source. 

We choose parameters similar to those used by Arrieta et al. in our simulations: a gaussian source of peak intensity $I_0 = 260 \; \mmol$, with a standard deviation $\sigma_I = 700\; \rm \mu m$. In the simple switch model, we simulate a change in phototactic sign at $I_c = I_0/2$, so that the algae exhibit positive phototaxis ($\alpha = 1)$ for $I \leq I_0/2$, and negative phototaxis otherwise. Algae are made to start at position $(x_0, y_0) = (500, 500) \; \rm \mu m$, at an angle of 200~degrees. The speed of the algae is $v_s = 50\; \rm \mu m/s$. We obtain loops for $\gamma = 10^{-5} \; \rm m^2\cdot\mu mol^{-1}$ and $s_T = 0.1$. This is a very different value of $\gamma$ from that used in our model. Yet, note that the algae in the experiments of Arrieta et al. were exposed to light for more than 10 minutes before being observed. It is likely that this induces adaptation, corresponding to a larger value of $s_T$ than that of our model, where cells were kept in the dark before the experiments. Taking another value of $s_T$ will affect the value of $\gamma$ for which loops are observed.

\subsection{Limits of the model}
\begin{figure*}[htb!]
  \centering
  \includegraphics[width=1\linewidth]{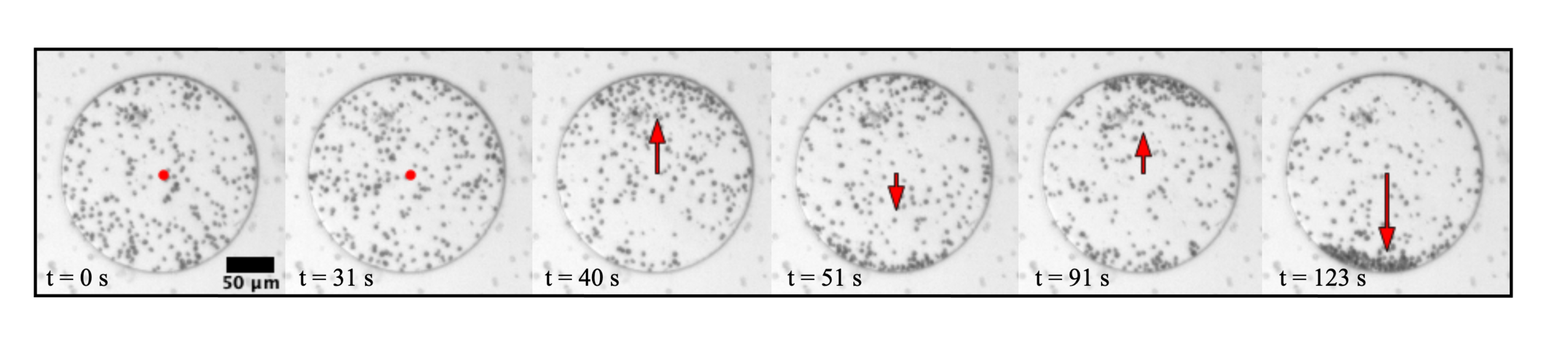}%
    \centering
    \caption{\chlamy\ enclosed in a well. A blue light stimulus of intensity $I = 2.0 \; \mmol$ is turned on at $t=30\; \rm s$. The stimulus comes from the upper side of the wells. In response to the stimulus, the algae go back-and-forth twice in the well.}
    \label{fig:oscillation}
\end{figure*}

The model is extremely simple. It does not reproduce some very rare cases we observed, where the algae go back-and-forth twice in the well, see Supp. Fig.~\ref{fig:oscillation}.  Such a behavior could potentially be recovered by introducing another time scale in the model, responsible for adaptation of the algae, which would lead to a change in time of the threshold $s_T$.

\bibliographystyle{apsrev4-2}
\bibliography{biblio_supp}% Produces the bibliography via BibTeX.

\end{document}